\documentclass{article}

\usepackage{upgreek}
\usepackage[numbers]{natbib}
\usepackage{graphicx}

\begin{document}

\title{Tapered multi-core fiber for lensless endoscopes}
\author{
  Fatima El Moussawi$^{1,4}$\\
  \and
  Matthias Hofer$^{2,4}$\\
  \and
  Damien Labat$^{1}$\\
  \and
  Andy Cassez$^{1}$\\
  \and
  Géraud Bouwmans$^{1}$\\
  \and
  Siddharth Sivankutty$^{1}$\\
  \and
  Rosa Cossart$^{3}$\\
  \and
  Olivier Vanvincq$^{1}$\\
  \and
  Hervé Rigneault$^{2}$\\
  \and
  Esben Ravn Andresen$^{1,*}$\\
}

\maketitle

\noindent
$^{1}$ Univ. Lille, CNRS, UMR 8523---PhLAM---Physique des Lasers, Atomes et Mol\'{e}cules, F-59000 Lille, France. \\
$^{2}$ Aix-Marseille Univ., CNRS, Centrale Marseille, Institut Fresnel, Marseille, France. \\
$^{3}$ INSERM, Aix-Marseille Univ., CNRS, INMED, Marseille, France. \\
$^{4}$ Contributed equally to this work. \\
$^{*}$ \texttt{esben.andresen@univ-lille.fr}

\section{abstract}
We present a novel fiber-optic component, a "tapered multi-core fiber (MCF)", designed for integration into ultra-miniaturized endoscopes for minimally invasive two-photon point-scanning imaging and to address the power delivery issue that has faced MCF based lensless endoscopes. 
With it we achieve experimentally a factor 6.0 increase in two-photon signal yield while keeping the ability to point-scan by the memory effect, and a factor 8.9 sacrificing the memory effect.
To reach this optimal design we first develop and validate a fast numerical model capable of predicting the essential properties of an arbitrarily tapered MCF from its structural parameters. We then use this model to identify the tapered MCF design parameters that result in a chosen set of target properties (point-spread function, delivered power, presence or absence of memory effect). We fabricate the identified target designs by stack-and-draw and post-processing on a CO$_{2}$ laser-based glass processing and splicing system. Finally we demonstrate the performance gain of the fabricated tapered MCFs in two-photon imaging when used in a lensless endoscope system. Our results show that tailoring of the taper profile brings new degrees of freedom that can be efficiently exploited for lensless endoscopes. 

\section{Introduction}
The term "lens-less endoscope" generally refers to an optical fiber based flexible imaging endoscope that reduces the diameter of the probe to the diameter of the fiber itself, typically 100--200~$\upmu$m; and where light at the distal end (closest to the sample) is controlled by wave front shaping at the proximal side (opposite to the sample) \cite{PsaltisOPN2016, andresen2016ultrathin}.
The lens-less endoscope is a promising ultra-miniaturized imaging tool which may enable minimally invasive and high-resolution observation of neuronal activity in-vivo inside deep brain areas \cite{TurtaevLSA2018, Vasquez-LopezLSA2018, OhayonBOE2018}. The interest of this miniaturized endoscope stems from its ability to allow new functionalities because light source and detectors are remote as well as the light weight and flexibility of the optical fibers which constitute here the main part of the imaging system. These properties give the lens-less endoscope the ability to be fixed for instance onto a mouse’s head allowing its free movement while studying neuronal activity. Also, it could reduce space constraints to allow fixing several probes simultaneously and thus the ability to study functional connectivity of neurons in two distant brain regions. The initial reports of lens-less endoscopy based on a bundle of single-mode fibers (SMFs) published in Ref.~\citenum{thompson2011adaptive}, and the first demonstrations of microscopic imaging of objects through multi-mode fiber which did not require any optical elements between the fiber and the object were published in Refs.~\citenum{vcivzmar2011shaping,choi2012scanner,papadopoulos2012focusing}. However, also multicore fiber (MCF) presents great potential as imaging wave guides due to its particular merits \cite{andresen2016ultrathin}. In this paper, we focus on MCF because it is more readily compatible with two-photon imaging modalities \cite{andresen2013two, andresen2016ultrathin, KimIEEEJSTQE2016} widely used in bio-medical microscopy. 

\begin{figure}[htbp]
  \centering
  \includegraphics[width=\textwidth]{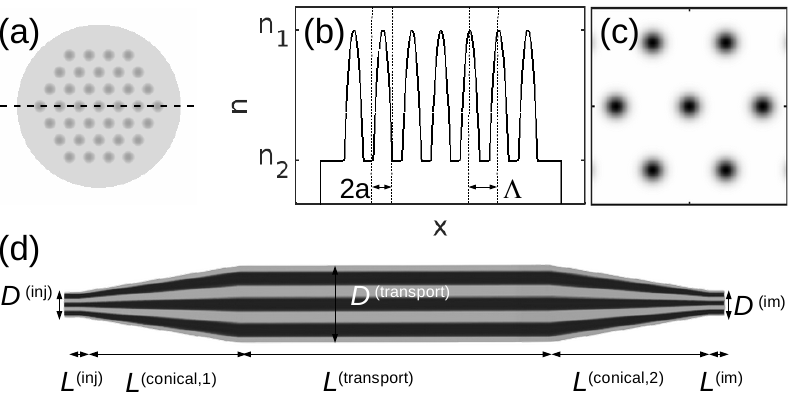}
\caption{Sketch of tapered MCF (shown with reduced core count, for compactness). 
(a) Structure of MCF with triangular core layout.
(b) Refractive index profile along the dashed line in (a). $a$, core diameter. $\Lambda$, pitch. $n_{1}$ Core refractive index. $n_{2}$ Cladding refractive index.
(c) Example image of intensity emerging from the distal end facet of the tapered MCF showing the fundamental mode of seven cores.
(d) Taper profile of the tapered MCF. From the left, 
injection segment of constant diameter $D^{(\mathrm{inj})}$ and length $L^{(\mathrm{inj})}$; 
first conical segment of length $L^{(\mathrm{conical,1})}$; 
transport segment of diameter $D^{(\mathrm{transport})}$ and length $L^{(\mathrm{transport})}$; second conical segment of length $L^{(\mathrm{conical,2})}$ ; 
imaging segment of diameter $D^{(\mathrm{im})}$ and length $L^{(\mathrm{im})}$.
\label{fig:sketch_doubletapered_MCF}}
\end{figure}

In a two-photon lens-less endoscope, the MCF must perform two tasks: (i) Imaging; and (ii) transport. For the imaging task the modes at the distal end face (closest to the sample)  of the MCF must lend themselves to coherent combination to intense foci for point-scanning imaging---best results have been obtained with dense core layouts with down to 3.2~$\upmu$m core separation as in Ref.~\citenum{conkey2016lensless}; while for the transport task, the main section of the MCF must be able to transport an ultra-short pulse distributed over all cores without deforming it through dispersion or changes induced by twists and bends---for which a sparse core layout with around 12--15~$\upmu$m core separation is better suited as in Refs.~\citenum{AndresenOL2013, andresen2013two, TsvirkunOptica2019}. 
In this paper we aim to realize a MCF comprising three segments comprising "injection", "transport" and "imaging" segments joined by conical tapers as sketched in Fig.~\ref{fig:sketch_doubletapered_MCF}. We will refer to this object with the shorthand "tapered MCF". Our aim is to reconcile the conflicting demands for both dense and sparse core layout in a tapered MCF in which the properties of injection, transport, and imaging segments are decoupled.

In this paper we report the design, fabrication, and application of a tapered MCF optimized for two-photon imaging with pulsed 920~nm excitation. 
The paper is organized as follows. In Sec.~\ref{sec:Results} we provide the main results of this work which we further discuss in Sec.~\ref{sec:Discussion}. Detailed Methods are provided in Supplement 1 and referenced when appropriate. The presented Data is available as Dataset 1, Ref.~\citenum{Dataset1}.
Specifically, in Sec.~\ref{subsec:Coupled_mode_theory_model} we present a fast numerical model capable of predicting the essential properties of an arbitrarily tapered MCF from its structural parameters. In Sec.~\ref{subsec:Target_properties_and_identification_of_target_designs} we use this model to identify the design parameters that result in a chosen set of target properties. In Sec.~\ref{subsec:Fabrication_and_characterization_of_target_designs} we report the fabrication and post-processing of the target designs as well as their optical characterization and validation. Finally, in Sec.~\ref{subsec:Two_photon_imaging_lensless_endoscope} we demonstrate the application of the fabricated tapered MCFs in a two-photon lensless endoscope system and quantify the gains in imaging performance.

\section{Results}
\label{sec:Results}

\subsection{Coupled-mode theory model}
\label{subsec:Coupled_mode_theory_model}

To assist in the identification of the optimal design parameters, we developed a code, based on coupled-mode theory (CMT) and perturbation theory, which we refer to as the "CMT model", and which is available as part of Dataset 1, Ref.~\citenum{Dataset1}. 
The CMT model considers the tapered MCF as a concatenation of uniform segments and requires the structural parameters of these segments. It returns the effective indices of the guided modes as well as its transmission matrix (TM) ${\mathbf{H}}$ which it calculates as the product of segment TMs: $\mathbf{H}$~=~$\prod_{i} {\mathbf{H}}_{i}$ (Suppl. Doc. Sec.~A).
The TM $\mathbf{H}$ is expressed in the basis of the fundamental modes of the $N$ cores and is of dimension $N \times N$ (higher-order modes---if present---are simply left out). 
$\mathbf{H}$ alone allows to predict the cross-talk (XT, a measure of the energy exchanged among cores) in terms of the $N \times N$ "cross-talk matrix" (sometimes called the "power transfer matrix") $\mathbf{X}$ here defined as
\begin{equation}
  \mathbf{X} = \langle | \mathbf{H} |^{2} \rangle _{\lambda}
\end{equation}
where the average over $\lambda$ is the average over a sufficiently large number of calculated TMs for different wavelengths (cf the experimental measurement with a filtered---about 10~nm spectral width---incoherent source);
and the scalar "average total cross-talk" $X_{\mathrm{ave}}$
\begin{equation}
  X_{\mathrm{ave}} = \frac{1}{N} \sum_{i} (\sum_{j \ne i} X_{ij} ) .
\end{equation}

\begin{figure}[htbp]
  \centering
   \includegraphics[width=\textwidth]{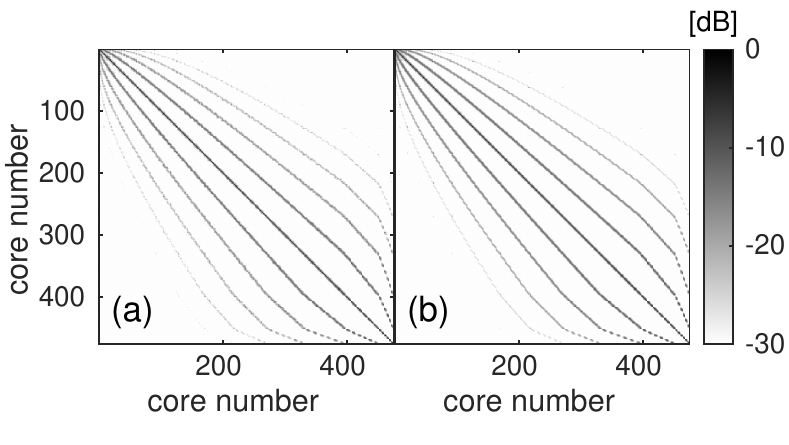}
  \caption{XT matrix $\mathbf{X}$ of a uniform MCF segment computed by (a) CMT model; and (b) FE-BPM model at 800~nm. Core numbering convention, the center core is number 1, the six cores in the first hexagonal ring are 2--7, etc. Simulation parameters: $L$~=~5~mm; $N$~=~475; $\Lambda$~=~7.5~$\upmu$m; $a$~=~1.4~$\upmu$m.}
  \label{fig:CMT-FEM}
\end{figure}

In parallel with the CMT model, we developed a model based on the finite-element beam propagation method (FE-BPM) \cite{rahman2013finite, koshiba1996wide} (Suppl. Doc. Sec.~B) which we refer to as "the FE-BPM model". Unlike CMT, FE-BPM makes no prior assumptions on the modes of the MCF and so is generally the more accurate of the two. It is however much slower. Here, we have used the FE-BPM model as a "gold standard" against which we have compared the results of the CMT model. 
In Fig.~\ref{fig:CMT-FEM}(a) and \ref{fig:CMT-FEM}(b) we present XT matrices for a $N$~=~475 core MCF as predicted by the CMT model and the FE-BPM model respectively which look qualitatively and quantitatively similar. The XT preferentially occurs between neighbor cores. This is not easy to see from the Figure because the core numbering convention influences the appearance of the Figure. Typically, in the modelled case there is significant XT ($X_{ij} > -30$~dB) between cores up to 5$\Lambda$ apart. From this comparison we establish the accuracy of the CMT model up to average total cross-talk of at least $X_{\mathrm{ave}}$~=~-2~dB and down to effective indices 0.8e-3 above the cladding index.  

The CMT model also returns the field of the fundamental core mode vs the radial coordinate; it can be interpolated on the cartesian "distal pixel basis" $\{ x^{(\mathrm{dist})}_j,y^{(\mathrm{dist})}_j \}, 1 \le j \le N_{\mathrm{pix}}$; centering $N$ copies on the core positions $(x^{(\mathrm{core})}_{j}, y^{(\mathrm{core})}_{j}), 1 \le j \le N$ ; each of these can be numerically propagated (assuming free-space propagation) to the far-field by a two-dimensional discrete Fourier transform, the resulting fields are expressed in the (angular) "far-field pixel basis" $\{ f^{(\mathrm{FF})}_{x,j},f^{(\mathrm{FF})}_{y,j} \}, 1 \le j \le N_{\mathrm{pix}}$; and finally re-arranged to a mode-to-far-field pixel basis change matrix $\mathbf{U}$ of dimensions $N \times N_{\mathrm{pix}}$. 
$\mathbf{U}$ together with $\mathbf{H}$ permit to calculate ${\mathbf{e}}^{(\mathrm{FF, pix})}$ of dimensions $N_{\mathrm{pix}} \times 1$ the field in the far-field of the MCF distal end face
\begin{equation}
  {\mathbf{e}}^{(\mathrm{FF, pix})} =  {\mathbf{U}}^{\dagger} \mathbf{H} {\mathbf{e}}^{(\mathrm{prox, mode})}
  \label{eq:useH}
\end{equation}
where ${\mathbf{e}}^{(\mathrm{prox, mode})}$ is $N \times 1$ and contains elements representing the complex modal field amplitudes injected into the proximal end of the MCF. It is from the vector ${\mathbf{e}}^{(\mathrm{FF, pix})}$ that the properties of Strehl ratio (a measure of the intensity in the distal focus relative to the total intensity) and the memory effect (a measure of input-output angular correlations allowing distal point-scanning by proximal angular scanning) \cite{freund1988memory} can be inferred. 
${\mathbf{e}}^{(\mathrm{FF, pix})}$ can also be numerically propagated to any intermediate plane with an angular spectrum propagator if required.
The elements of the $N \times 1$ vector ${\mathbf{e}}^{(\mathrm{prox},\mathrm{mode},j_{0})}$ are the optimal input modal amplitudes to obtain an optimal focus on far-field pixel $j_{0}$ and is found as the conjugate transpose of the $j_{0}$'th row of $\mathbf{U}^{\dagger}\mathbf{H}$.
${\mathbf{e}}^{(\mathrm{FF, pix, j_{0}})}$ is the resulting complex field optimally focused on far-field pixel $j_{0}$ which is found by applying Eq.~(\ref{eq:useH}). 
The Strehl ratio $S$ can then be defined as the ratio of the energy in the focus on far-field pixel $j_{0}$ to the total energy in the field, for example by plotting the norm-square of ${\mathbf{e}}^{(\mathrm{FF, pix, j_{0}})}$ versus $\{ f^{(\mathrm{FF})}_{x,j},f^{(\mathrm{FF})}_{y,j} \}$ and integrating over the focus area and then over the whole field.
The so-called "memory effect" is the effect by which the direction (phase ramp) of the output field changes by the same amount as the direction (phase ramp) of an input field \cite{freund1988memory}. In presence of the memory effect, it is therefore possible to displace a distal focus by simply adding a phase ramp to the proximal field.
${\mathbf{p}}(f^{(0)}_{x}, f^{(0)}_{y})$ is $N \times 1$ and its elements are the [complex exponentials of] additional proximal modal phases required to move the focus in the far-field by $(f^{(0)}_{x},f^{(0)}_{y})$ using the memory effect:
\begin{equation}
  \mathbf{p}(f^{(0)}_{x},f^{(0)}_{y}) : p_{k}(f^{(0)}_{x},f^{(0)}_{y}) = \mathrm{exp} (i 2 \pi f^{(0)}_{x} x_{k}^{(\mathrm{core})} + i 2 \pi f^{(0)}_{y} y_{k}^{(\mathrm{core})} )
  \label{eq:phaseramp}
\end{equation}
The memory effect can then be quantified as a two-dimensional curve $M(f^{(0)}_{x},f^{(0)}_{y})$ which is a ratio, the denominator containing the intensity of the field "optimally focused on far-field pixel $j_{1}$" ${\mathbf{e}}^{(\mathrm{FF, pix, j_{1}})}$; 
and the numerator containing the intensity of the field "initially optimally focused on far-field pixel $j_{0}$ then shifted to far-field pixel $j_{1}$ using the memory effect":
\begin{equation}
  {\mathbf{e}}^{(\mathrm{FF, pix}, j_{0}\rightarrow j_{1})} = {\mathbf{U}}^{\dagger} \mathbf{H} [ {\mathbf{h}}^{(j_{0})} \circ {\mathbf{p}}(f^{(0)}_{x}, f^{(0)}_{y}) ]
\end{equation}
where "$\circ$" signifies Hadamard product (element-by-element multiplication).
$M$ is constant =~1 for "full" memory effect, and =~0 everywhere except the origin for "no" memory effect. 

As outlined above, the quantities $X_{\mathrm{ave}}$, $S$, and $M$ represent properties of tapered MCF that can be predicted by the CMT model from the tapered MCFs structural properties. Inversely, the CMT model can be used to identify the tapered MCF designs that result in a chosen set of properties. 

\subsection{Target properties and identification of target designs}
\label{subsec:Target_properties_and_identification_of_target_designs}

The objective is to identify the design parameters cf Fig.~\ref{fig:sketch_doubletapered_MCF}(d) of a tapered MCF with triangular core layout with suitable properties for two-photon lensless imaging with a pulsed 920~nm excitation laser---the method would also work for any other core layout one has the capacity to fabricate. 

\begin{figure}[htbp]
  \centering
  \includegraphics{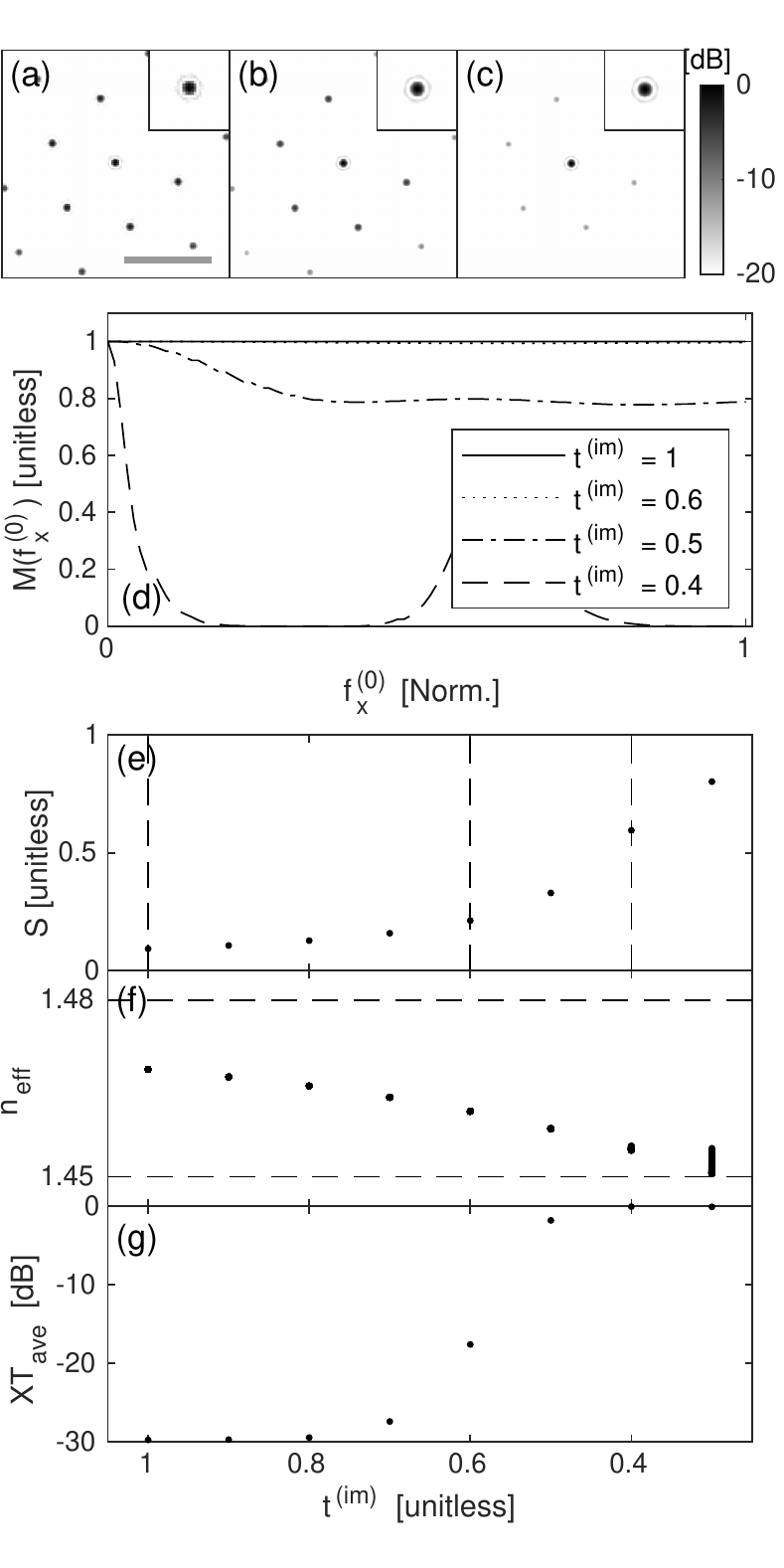}
  \caption{Properties of tapered MCF predicted by CMT model. 
  (a)-(c) PSF for $t^{(\mathrm{im})}$~=~1.0, 0.6, and 0.4 in a plane $Z$~=~$Z_{0}\cdot t^{(\mathrm{im})}$.
  Each image is normalized to its maximum intensity.
  Scale bar, 50~$\upmu$m.
  Insets, 5$\times$ zoom on the central spot.
  (d) Memory effect curves. Horizontal axis normalized to half the distance between replica spots.
  (e) Strehl ratio.
  (f) Effective indices of supermodes. 
  (g) Average XT. 
  Unmentioned parameters are as in Tab.~\ref{tab:taperedMCFsamples}.
  }
  \label{fig:Identification}
\end{figure}

As a first step, we want to minimize inter-core group delay dispersion in the transport segment cf Fig.~\ref{fig:sketch_doubletapered_MCF}(d) (center) which is debilitating for the short-pulsed excitation that we aim for. This dispersion arises from local variations among cores. We identify the rule-of-thumb that inter-core group delay dispersion decreases with increasing $V$-parameter. This holds for all the hypothesized physical origins that we considered  (\cite{roper2015minimizing, roper2015advances, andresen2015measurement} and Suppl. Doc. Sec.~D). We therefore choose to constrain the $V$-parameter to around 5.06, the second cutoff of a parabolic-index core, making the core bi-mode. 
To avoid exciting the higher-order mode in the transport segment, we decide to add an "injection" segment separated cf Fig.~\ref{fig:sketch_doubletapered_MCF}(d) (left) from the transport segment by a conical taper whose dimensions are scaled by the factor $t^{(\mathrm{inj})}$~=~0.6 resulting in $V$~=~3.0 which is comfortably below the single-mode cutoff ($V_{\mathrm{cutoff}}$~=~3.518). 
Next we want to fix the refractive index step, core size, and pitch in the transport segment with the aim of keeping $X_{\mathrm{ave}}$ very low, say, below -25~dB in one metre of MCF. To this end, the refractive index step should generally be chosen as high as practically possible, $\Delta n$~=~30$\cdot$10$^{-3}$; to conserve $V$~=~5 the radius must then be chosen $a$~=~2.5~$\upmu$m. The only parameter left is the pitch $\Lambda$. We use the CMT model to predict that the chosen criterion on $XT_{\mathrm{ave}}$ is respected for $\Lambda~>~14~\upmu$m (predicted $X_{\mathrm{ave}}$~=~-62~dB per metre in the uniform segment, -26~dB in the injection segment with $t^{(\mathrm{inj})}$~=~0.6). 
At this point the number of cores $N$ can be fixed. It may be chosen arbitrarily, generally "the more the better" as the number of resolvable image pixels is proportional to $N$. Other than that $N$ only impacts on the outer diameter of the transport section, so it can be chosen so as to respect a chosen condition on the outer diameter. Here, we choose $N$~=~349, resulting in an outer diameter $D^{(\mathrm{transport})}$~=~330~$\upmu$m. 
Finally, and most importantly, we now want to choose the dimensions of the imaging segment cf Fig.~\ref{fig:sketch_doubletapered_MCF}(d) (right). All of its structural parameters are scaled by the factor $t^{(\mathrm{im})}$ relative to those of the transport segment. We are particularly interested in two values, a first one which maximizes the Strehl ratio under the condition that the "full" memory effect is conserved; and a second one which maximizes the Strehl ratio without any conditions on the memory effect.
But in this case a new condition must be respected, that core guidance is maintained, i.e. the effective indices of all modes must remain comfortably above the cladding index so as not to be too close to the single-mode cutoff. 
In Fig.~\ref{fig:Identification}(d) we present the curves of memory effect for different $t^{(\mathrm{im})}$; in Fig.~\ref{fig:Identification}(e) the Strehl ratio; in Fig.~\ref{fig:Identification}(f) the effective indices; and in Fig.~\ref{fig:Identification}(g) the average XT as functions of $t^{(\mathrm{im})}$ as predicted by the CMT model (the rest of the parameters are as in Tab.~\ref{tab:taperedMCFsamples}). Based on the following discussion we identify $t^{(\mathrm{im})}$~=~0.6 (maximum Strehl ratio while retaining full memory effect) and 0.4 (maximum Strehl ratio while retaining core guidance) respectively. 

Fig. \ref{fig:Identification}(a)-\ref{fig:Identification}(c) present the predicted point-spread functions (PSF) at distance $Z$~=~$Z_{0} \cdot t^{(\mathrm{im})}$ from the distal end of the MCFs with $Z_{0}$~=~500~$\upmu$m and $t^{(\mathrm{im})}$~=~1, 0.6, and 0.4 (\textit{per se} they are far-field images scaled by $x$~=~$f_{x} Z \lambda$, $y$~=~$f_{y} Z \lambda$). It comprises a central spot surrounded by a dimmer periodic replicas for untapered MCF ($t^{(\mathrm{im})}$~=~1). However, it becomes more singly-peaked, very beneficial for image quality, as the taper ratio decreases while the focal spot size which is represented by the full-width at half-maximum (FWHM) of the PSF is fairly constant equal to 2.4~$\upmu$m. From Fig.~\ref{fig:Identification}(f) which shows the predictions of effective indices it can be seen that at $t^{(\mathrm{im})}$~=~0.3 some of the indices drop below the cladding index (lower dashed line) making leaking to cladding modes inevitable. This is the reason the lowest value of $t^{(\mathrm{im})}$ retained as design parameter was 0.4 for which effective indices remain 2$\cdot$10$^{-3}$ above the cladding index. Another important consequence of tapering, of primordial importance for two-photon imaging, is the dramatic increase in the Strehl ratio with decreasing taper ratio as shown in Fig. \ref{fig:Identification}(e). 

\subsection{Fabrication and characterization of target designs}
\label{subsec:Fabrication_and_characterization_of_target_designs}
We fabricated a MCF to the specification requirement of the transport segment identified in Sec.~\ref{subsec:Target_properties_and_identification_of_target_designs} ($\Delta n$~=~30$\cdot 10^{-3}$; $\Lambda$~=~14~$\upmu$m; $a$~=~2.5~$\upmu$m; $N$~=~349) by the methods we used previously (Ref.~\citenum{AndresenOL2013} and Suppl. Doc. Sec.~C). 

Then we post-processed the uniform MCF to obtain the tapered MCF represented by Fig.~\ref{fig:sketch_doubletapered_MCF}(d) by tapering a short length at the terminals of the MCF using a CO$_{2}$ laser-based glass processing and splicing system (LZM-100 FUJIKURA) (Suppl. Doc. Sec.~E). The MCF is locally heated up with the CO$_{2}$ laser and simultaneously pulled lengthwise by stepper motors so as to obtain a fiber whose diameter varies longitudinally. This system provides extremely stable operation and allows control over a wide range of parameters to achieve the tapered MCFs precisely with the chosen design parameters Tab.~\ref{tab:taperedMCFsamples}.

\begin{table}[htbp]
    \centering
    \begin{tabular}{|c|c|c|c|}
        \hline
        Identifier & 0 & 1 & 2 \\
        \hline
        $D^{(\mathrm{inj})}$~[$\upmu$m]  & 196 & 194 & 196 \\
        $t^{(\mathrm{inj})}$  & 0.6 & 0.6 & 0.6 \\
        $L^{(\mathrm{inj})}$~[m]  & 0.005 & 0.005 & 0.005 \\
        $L^{(\mathrm{conical,1})}$~[m]  & 0.025 & 0.025 & 0.025 \\
        $D^{(\mathrm{transport})}$~[$\upmu$m]  & 330 & 330 & 330 \\
        $L^{(\mathrm{transport})}$~[m]  & 0.23 & 0.21 & 0.24 \\
        $L^{(\mathrm{conical,2})}$~[m]  & 0 & 0.05 & 0.05 \\
        $D^{(\mathrm{im})}$~[$\upmu$m]  & 330 & 199 & 135 \\
        $t^{(\mathrm{im})}$  & 1 & 0.6 & 0.4 \\
        $L^{(\mathrm{im})}$~[m]  & 0 & 0.005 & 0.005 \\
        \hline
    \end{tabular}
    \caption{ The physical dimensions of the fabricated tapered MCF samples.}
    \label{tab:taperedMCFsamples}
\end{table}

We characterized the optical properties of the fabricated tapered MCF using a filtered super-continuum laser source (Suppl. Sec.~F). We injected the focused laser beam into MCF cores in the injection segment successively, and imaged the emerging beam out of imaging segment on a camera. Consequently, we could measure the mode field diameter (MFD) of the visualized mode profile and confirm---for each $t^{(\mathrm{im})}$---the agreement with the predictions of the CMT model. Additionally we were able to measure the XT between excited core and other cores and thus build up $\mathbf{X}$ in the core basis which is presented in Fig.~\ref{fig:Expproperties}(f) and which agrees reasonably with the predicted values in Fig.~\ref{fig:Identification}(g). To confirm that the tapered MCFs were effectively single-mode with the choice of injection segment parameters, we tested a long MCF ($L^{(\mathrm{transport})}>~1$~m) with $t^{(\mathrm{inj})}$~=~0.6 and $t^{(\mathrm{im})}$~=~1. On the camera we visualized the intensity profile exiting one core and verified that it did not change while applying a perturbation on the fiber. This verifies that only the fundamental mode propagates through the transport segment despite it being bi-modal, thus the injection segment allows the selective excitation of only the fundamental mode of each core.

\subsection{Two-photon imaging lensless endoscope}
\label{subsec:Two_photon_imaging_lensless_endoscope}

\begin{figure}[htbp]
    \centering
    \includegraphics{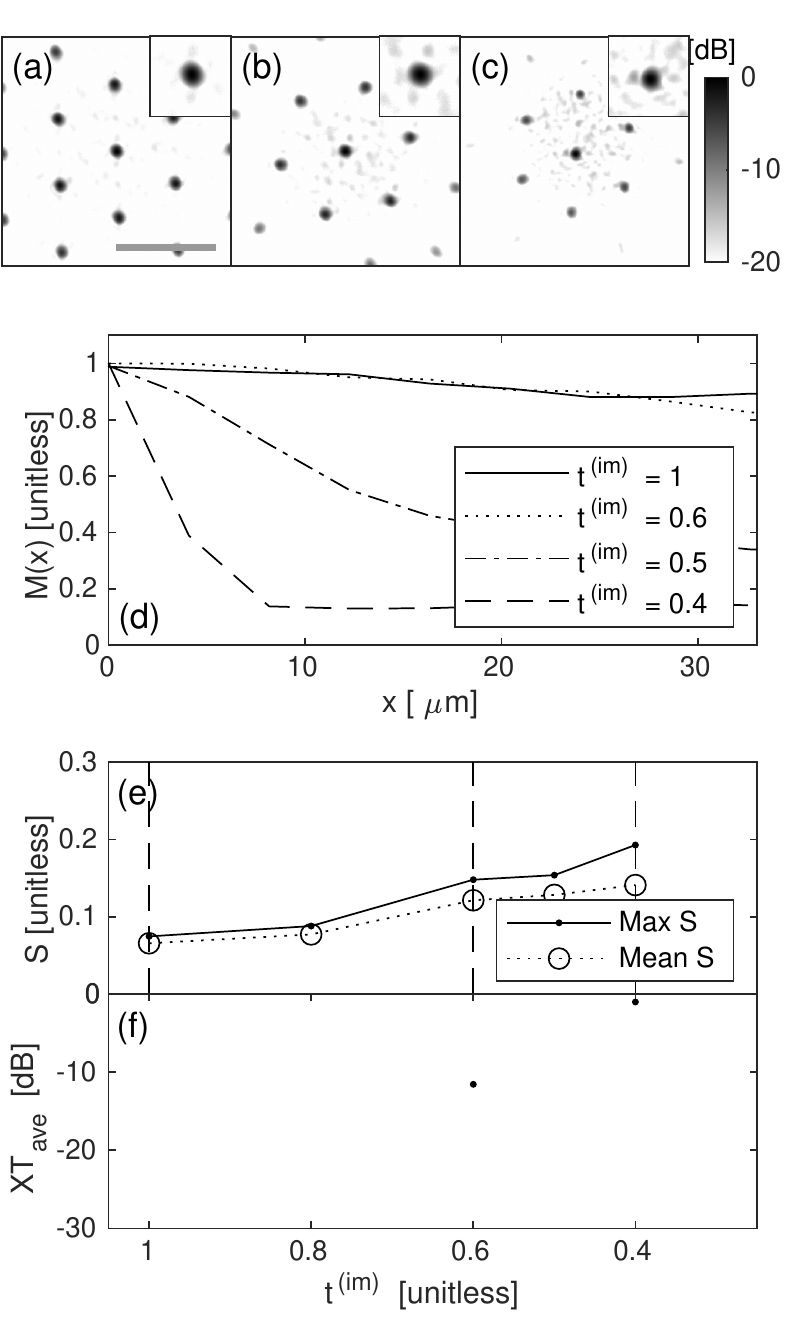}
    \caption{Experimentally observed properties of the tapered MCFs with parameters given in Tab.~\ref{tab:taperedMCFsamples}; the parameters of the two tapered MCFs with $t^{(\mathrm{im})}$~=~0.8 and 0.5 not appearing in the Table are very similar. (a)--(c) PSF for $t^{(\mathrm{im})}$~=~1, 0.6, and 0.4 in a plane $Z$~=~$Z_{0} \cdot t^{(\mathrm{im})}$~=~500, 300, and 200~$\upmu$m. Insets, 5$\times$ zoom on the central spot. Each image is normalized to its maximum intensity. Scale bar, 50~$\upmu$m.
    (d) Memory effect curves.
    (e) Strehl ratio. 
    (f) Average cross-talk.
    \label{fig:Expproperties}}
\end{figure}

As a first step towards implementing the tapered MCFs in a lensless endoscope system we measured the imaging properties of the fabricated tapered MCFs. The measurements were done using a pulsed fs-laser shaped by a spatial-light modulator (SLM) for injection while measuring the resulting intensity distributions in a plane a distance $Z$ from the distal end face. The SLM was programmed to generate the focus (Suppl. Doc. Sec.~G). The experimentally measured properties are presented in Fig.~\ref{fig:Expproperties} and organized in the same way as the Figure of predicted properties, Fig.~\ref{fig:Identification}. 
Figures~\ref{fig:Expproperties}(a)--\ref{fig:Expproperties}(c) show the PSFs with $t^{(\mathrm{im})}$~=~1, 0.6, and 0.4 and $Z$~=~$Z_{0} \cdot t^{(\mathrm{im})}$~=~500~$\upmu$m, 300~$\upmu$m, and 200~$\upmu$m. The measured FWHMs of the PSFs are 4.2, 4.0, and 3.6~$\upmu$m, slightly larger than predicted. The rest of the observed trends are in line with the predictions, that is, increasing intensity in the central focus and decreasing intensity in the satellite peaks as $t^{(\mathrm{im})}$ is decreased. 
This trend is quantified in the measured Strehl ratio, presented in Fig.~\ref{fig:Expproperties}(e), from a value of 7~\% at $t^{(\mathrm{im})}$~=~1, it increases to 15~\% at $t^{(\mathrm{im})}$~=~0.6 and reaches 19~\% at the smallest value of $t^{(\mathrm{im})}$~=~0.4. 
In terms of absolute power in focus these values correspond to 20, 43, and 54~mW.
The experimentally measured memory effect curves are presented in Fig.~\ref{fig:Expproperties}(d). 
The taper ratios $t^{(\mathrm{im})}$~=~1, 0.6, and 0.4 were identified as key values in Sec.~\ref{subsec:Target_properties_and_identification_of_target_designs}, the first assuring maximum memory effect, the second assuring the highest Strehl ratio while preserving significant memory effect, and the final assuring highest Strehl ratio regardless of memory effect. The Fig.~\ref{fig:Expproperties}(d) confirms these predictions, as the curves $t^{(\mathrm{im})}$~=~1 and 0.6 are coincident and almost constant equal to 1 testifying to an almost identical memory effect. On the other hand, the curve $t^{(\mathrm{im})}$~=~0.4 strongly peaked around $x$~=~0 testifies to an almost complete absence of memory effect but the highest Strehl ratio as seen from Fig.~\ref{fig:Expproperties}(e). 

\begin{figure}[htbp]
    \centering
    \includegraphics[width=\textwidth]{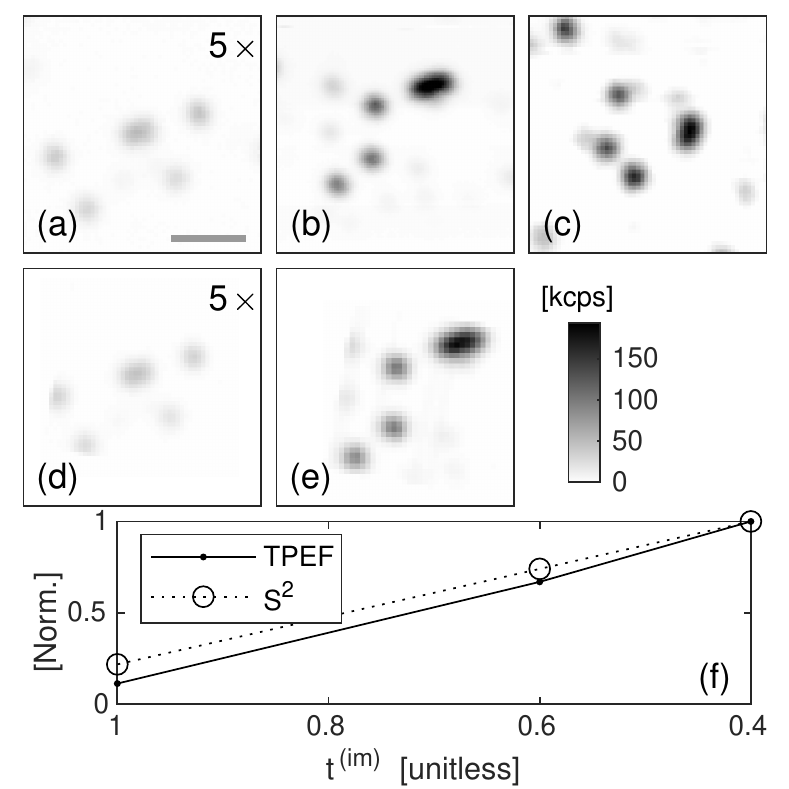}
    \caption{Two-photon excited fluorescence images of fluorescent beads with the lensless endoscope. 
    (a)-(c) Point-scanning with an optimal focus on each pixel (TM scan) and taper ratios (a) $t^{(\mathrm{im})}$~=~1 and $Z$~=~500~$\upmu$m; (b) $t^{(\mathrm{im})}$~=~0.6 and $Z$~=~300~$\upmu$m; and (c) $t^{(\mathrm{im})}$~=~0.4 and $Z$~=~200~$\upmu$m. 
    (d)-(e) Point-scanning with memory effect starting from an optimal focus on the central pixel (ME scan) and taper ratios (d) $t^{(\mathrm{im})}$~=~1; and (e) $t^{(\mathrm{im})}$~=~0.6.
    Spatial scales are identical for all the images. Intensities in (a) and (d) have been multiplied by 5 for visualization. Scale bar, 10~$\mu$m.
    (f) Evolution of two-photon signal with taper ratio. (dots) Observed two-photon signals; and (circles) the square of the observed Strehl ratios. }
    \label{fig:TPEF}
\end{figure}

The formation of a two-photon fluorescence image in a lensless endoscopy system requires a single-point detector to acquire the two-photon fluorescence generated by the sample; and a raster scanning scheme where the focus is scanned across two dimensions. This can be realized in two different ways, either (i - "TM scan") the phases of the elements of a pre-measured TM can be converted to SLM masks which scan the focus across the sample when displayed sequentially on the SLM; or (ii - "memory effect (ME) scan") a focus initially on a center pixel can be displaced by the memory effect and scanned across the sample, this is done through the addition of a sequence of suitable phase ramps in a conjugate plane of the proximal MCF end face (Suppl. Sec.~G). When possible, the ME scan method has the dual advantage of not requiring knowledge of a full TM; and the possibility to use fast galvo mirrors to generate the phase ramps (virtually unbound image acquisition time \cite{AndresenOL2013}). 
With the different tapered MCFs in our lensless endoscope system, we performed two-photon imaging with both the TM scan method and the ME scan method. For simplicity and pragmatism, here, we used the SLM for both (so we are unable to demonstrate the speed advantage). 
For these proof-of-principle demonstrations, we used yellow-green fluorescent beads as a sample whose emission spectrum is similar to that of yellow fluorescent proteins commonly expressed e.g. in mice neurons during in-vivo exploration of neuronal activity. 
The acquired two-photon images are presented in Fig.~\ref{fig:TPEF}. 
The upper row Figs.~\ref{fig:TPEF}(a)-\ref{fig:TPEF}(c) show two-photon images acquired using the TM scan method with $t^{(\mathrm{im})}$~=~1, 0.6, and 0.4. 
The second row Figs.~\ref{fig:TPEF}(d), \ref{fig:TPEF}(e) show two-photon images acquired using the ME scan method with $t^{(\mathrm{im})}$~=~1 and 0.6. The comparison to Figs.~\ref{fig:TPEF}(a), \ref{fig:TPEF}(b) provides visual evidence that the memory effect is indeed retained at these values, as predicted. No image with the ME scan is shown for $t^{(\mathrm{im})}$~=~0.4 because the ME is absent. 
All images are on the same intensity scale which allows to appreciate the increase in two-photon signal with decreasing $t^{(\mathrm{im})}$. It can be seen that the signal from single, isolated beads increases appreciably from Fig.~\ref{fig:TPEF}(a) over \ref{fig:TPEF}(b) to \ref{fig:TPEF}(c). 
This two-photon signal increase is quantified in Fig.~\ref{fig:TPEF}(f). The points represent a mean over two-photon signal measured on several single, isolated beads. The relative increases are a factor 6.0  and a factor 8.9 for $t^{(\mathrm{im})}$~=~0.6 and 0.4. 
The two-photon signal increase is expected to be proportional to the square of the Strehl ratio, represented by circles. And indeed the two curves are coincident to a high degree.

\section{Discussion}
\label{sec:Discussion}

Initially, we remark that for the experimental two-photon imaging, only 169 out of the $N$~=~349 cores of the tapered MCFs were employed simultaneously. 
An experimental constraint limiting the number of cores that can be used simultaneously is the number of pixels on the SLM. A 800$\times$600 pixel SLM can be divided into 169 40$\times$40 pixel segments which must control not only the phase but also the focusing of light injected into a core (Suppl. Sec.~G), and the latter is compromised as the number of pixels per segment decreases. So, while not a hard limit, 169 cores was the compromise we chose. Workarounds are to use one of the newer SLM generations with higher pixel count; or using the SLM in conjunction with a passive elements like a micro-lens array to take care of the focusing.
The modelling was limited to the same number of cores ($N$~=~169) in order to have directly comparable results. 

Next, we remark that the CMT model has not correctly predicted the experimentally measured Strehl ratios [Fig.~\ref{fig:Identification}(e) compared to Fig.~\ref{fig:Expproperties}(e)]. For $t^{(\mathrm{im})}$~=~0.6 the prediction is $S$~=~21~\% versus the observed $S$~=~15~\%. 
This discrepancy can be accounted for by two factors not included in the model.
(i) Polarization is not accounted for in the CMT model, but the fabricated tapered MCF has circular ie non-polarization-maintaining cores. In the experiment a polarizer selects one of the polarization components, each core thus contributes a random amplitude to the image on the camera. 
The effect can be modelled by multiplying each element of ${\mathbf{e}}^{(\mathrm{prox}, \mathrm{mode}, j_{0})}$ by a random number between zero and one before applying Eq.~(\ref{eq:useH}), and the consequence is a drop of $S$ down to 18~\%. 
(ii) The model calculates the Strehl ratio in the far-field whereas the experiments were done in intermediate planes $Z$~=~500, 300, and 200~$\upmu$m. The predicted Strehl ratio should therefore be taken as upper bounds. If we numerically propagate the predicted far-field to the intermediate plane, this effect alone leads to a drop of $S$ down to 16.5~\%. Both effects together result in a drop of $S$ to 12~\%. These effects thus seem to fully explain the discrepancy.
For $t^{(\mathrm{im})}$~=~0.4 the discrepancy is more flagrant, the prediction is $S$~=~59~\% versus the observed $S$~=~19~\%. Here; effects (i) and (ii) combined would only result in a drop of $S$ down to 42~\%. 
It is possible that at this small value of $t^{(\mathrm{im})}$ the effect of leakage from the cores into the cladding becomes important. This would signify that the CMT model is at the limit of or even beyond its regime of validity for $t^{(\mathrm{im})}$~=~0.4. Beyond this point one would have to resort to FE-BPM modelling to assure accurate results (using, for example, the open source FE-BPM tool in Ref.~\citenum{VeettikazhyOE2021}).

Inter-core group delay dispersion (GDD) \cite{andresen2015measurement} is also not accounted for in the CMT model. Non-zero inter-core group delay dispersion has the consequence of reducing the coherence between the fields emerging from different cores (since we are using ultra-short pulses) and hence the Strehl ratio. 
We have sampled inter-core GDD in some pairs of cores which indicate that the distribution of inter-core GDD is narrower than the duration (150~fs) of the ultra-short laser pulse. This is not expected to be a main source of the discrepancy in Strehl ratios and the above discussion corroborates this.

We must remark that the lensless endoscope system used to generate Figs.~\ref{fig:Expproperties} and \ref{fig:TPEF} employed a detector located at the distal end, after both MCF and sample, rather than at the proximal end as it would in a "true" endoscopic configuration, detecting fluorescence collected in the backward direction through the MCF. 
In a previous paper we did demonstrate how a double-clad MCF allows for efficient fluorescence collection and thus to work with a proximal detector only \cite{andresen2013two}, where the double-cladding was a ring of air holes around the $N$ cores. However we were unable to use the same kind of double cladding in the present work because the post-processing on the glass processing and splicing system would have collapsed the air holes. 
In the near term our aim is to realize versions of the tapered MCF presented above with double cladding. We envisage that coating the tapered MCFs with low-index polymer as a final step will achieve this effect. 

One of the biggest challenges facing lensless endoscopes---be they multi-mode fiber-based or MCF-based---is the resilience to bending or lack thereof. Conformational changes of a multi-mode fiber or MCF generally alters its TM significantly, meaning that re-performing the initial calibration steps is mandatory after each conformational change, which becomes unfeasible even for slow conformational changes. 
The tapered MCFs presented above are also subject to this problem---even though we previously showed that bending a MCF results in a predictable phase ramp which only shifts the focus without degrading it \cite{TsvirkunOE2017}. 
In a more recent paper \cite{TsvirkunOptica2019} we reported the fabrication of a helically-twisted MCF and demonstrated that it is invariant to conformational changes in a fairly wide range of operational conditions when used in a lensless endoscope system. 
The fabrication method of the tapered MCFs outlined in the present paper takes as its starting point a standard (ie not helically twisted) MCF. But there would be no conceptual difficulties starting from a helically-twisted MCF instead. As such the pursuit of conformationally-invariant tapered MCFs also represents a near-term aim. 

\section{Conclusion}
We have presented a novel tapered MCF for optimal two-photon lensless endoscopy by point-scanning imaging. Our key result is that optimized tapered MCF can increase two-photon yield by a factor 6.0 while still keeping the ability to point-scan by the memory effect, and up to a factor 8.9 while sacrificing the memory effect. These results are decoupled from the transport segment of the tapered MCF with low XT which can be metres in length without altering the results. We have outlined the entire procedure of conception, design, fabrication, and application in a lensless endoscopy system of this novel fiber-optic component. The procedure, namely the step involving the CMT model, is generic and applicable to other ranges of properties, other MCF core layouts e.g. a Fermat golden spiral layout \cite{SivankuttyOL2018}, wavelength ranges, etc. We conclude that tailoring of the taper profile is a degree of freedom that can be efficiently exploited for ameliorating MCF based lensless endoscopes and we have shown here that it contributes to solving the major power delivery issue that MCF based lensless endoscopes \cite{SivankuttyOL2021}. More generally a tapered multi-core fiber is a valuable add-on into the lensless endoscope toolbox to design, optimize and fabricate ultra-miniaturized endoscope tailored to a broad range of applications.

\section{Backmatter}

\textbf{Funding}\\
Agence Nationale de la Recherche (ANR-11-IDEX-0001-02,  ANR-20-CE19-0028, ANR-21-ESRS-0002 IDEC, ANR-21-ESRE-0003 CIRCUITPHOTONICS); 
INSERM (18CP128-00, PC201508); 
Centre National de la Recherche Scientifique,  Aix-Marseille Université (A-M-AAP-ID-17-13-170228-15.22-RIGNEAULT); 
National Institute of Health (NIH R21 EY029406-01).

\textbf{Acknowledgments}\\
Parts of this work were developed at IRCICA (USR CNRS 3380, https://ircica.univ-lille.fr/) using FiberTech Lille facilities (https://fibertech.univ-lille.fr/en/). 
This work was supported by the French Ministry of Higher Education and Research, the "Hauts de France" Regional Council, the European Regional Development fund (ERDF) through the CPER "Photonics for Society".

\textbf{Disclosures}\\
The authors declare no conflicts of interest.

\textbf{Supplemental document}\\
See Supplement 1 for supporting content.

\end{document}


\title{Supplementary Document: Tapered multi-core fiber for lensless endoscopes}
\author{
  Fatima El Moussawi$^{1,4}$\\
  \and
  Matthias Hofer$^{2,4}$\\
  \and
  Damien Labat$^{1}$\\
  \and
  Andy Cassez$^{1}$\\
  \and
  Géraud Bouwmans$^{1}$\\
  \and
  Siddharth Sivankutty$^{1}$\\
  \and
  Rosa Cossart$^{3}$\\
  \and
  Olivier Vanvincq$^{1}$\\
  \and
  Hervé Rigneault$^{2}$\\
  \and
  Esben Ravn Andresen$^{1,*}$\\
}

\maketitle

\noindent
$^{1}$ Univ. Lille, CNRS, UMR 8523---PhLAM---Physique des Lasers, Atomes et Mol\'{e}cules, F-59000 Lille, France. \\
$^{2}$ Aix-Marseille Univ., CNRS, Centrale Marseille, Institut Fresnel, Marseille, France. \\
$^{3}$ INSERM, Aix-Marseille Univ., CNRS, INMED, Marseille, France. \\
$^{4}$ Contributed equally to this work. \\
$^{*}$ \texttt{esben.andresen@univ-lille.fr} 

\appendix

\section{Coupled-mode theory model}
\label{SI:sec:Coupled_mode_theory_model}
In this section we present the coupled-mode theory (CMT) method to model a tapered multi-core optical fiber (MCF). A Matlab function implementing the essential of the CMT model is available in Dataset 1, Ref.~\citenum{Dataset1}, its input and output arguments are further detailed in Tab.~\ref{SI:tab:variables}.
We start out with a MCF containing $N$ cores arranged in a multi-ring hexagonal lattice with pitch $\Lambda$ within a diameter $D$. The model is also compatible with arbitrary core positions. We begin our analysis by assuming that each core is cylindrical with radius $a$, and single moded, i.e. it only supports the LP01 mode with a propagation constant $\beta_0$. When waveguides are brought into close proximity, the modes will couple to each other as a result of the overlap of the evanescent fields. Thus, the spatial dependence of one mode amplitude will be modified by the existence of the other. In the case of weak coupling, the coupled mode equations are be of the form:

\begin{eqnarray}
\frac{\mathrm{d} A_{p}}{\mathrm{d} z} + i \beta_{0} A_p = i \sum_{q=1}^{N} C_{pq} A_q
\label{eqn:A}
\end{eqnarray}

\noindent where $A_{p}$ ($A_{q}$) are the complex amplitudes of mode $p$ ($q$); and $C_{pq}$ is the coupling coefficient between modes $p$ and $q$ and which can be obtained using CMT \cite{snyder2012optical}

\begin{eqnarray}
	C_{pq} = \frac{k_0}{2 n_1} \int_{S_\infty} [n_{\mathrm{pert}}^2(x,y)-{n}^2(x,y)]~ \psi_{p}~ \psi_{q}~ \mathrm{d}S
	\label{eqn:Couplingcoef}
\end{eqnarray}

\noindent where  $n_{\mathrm{pert}}(x,y)$ and $n(x,y)$ are the refractive index profiles of perturbed and the non-perturbed waveguides respectively, $\psi_{p}$ and $\psi_{q}$ are the normalized modal fields, and  $S_\infty$ is the infinite cross-section.  

We assume that an unperturbed core has a refractive index $n(x,y)$ as an isolated core. And the perturbed refractive index $n_{\mathrm{pert}}(x,y)$ is the sum of the refractive indices of cores $p$ and $q$. $n_{\mathrm{pert}}^{2}(x,y) - n^{2}(x,y)$ is the perturbation. Therefore, the coupling between each pair of cores is computed from the overlap between their fields, considering the perturbation in the index profile Eq.~(\ref{eqn:Couplingcoef}). Assuming that the mode propagation equation in Eq.~(\ref{eqn:A}) has the a solution of the form $ A_p = A_{p0} \mathrm{exp}(i \beta z) $, where $A_{p0}$ is slowly varying with z, we get a slowly varying function with the derivative of the perturbation function, obtaining an equation in the form of an eigenvalue problem,

\begin{eqnarray}
\Delta{\beta} A_{p} =  \sum_{q} C_{pq} A_{q} .
\end{eqnarray}
 
From this eigenvalue equation we will be able to determine the eigenvalues that are represented by the difference of the perturbed and non-perturbed propagation constants $\Delta{\beta}$. And as we are studying the coupling between $N$ cores in a MCF, we represent the coupling equation in matricial form. Thus the coupling coefficients matrix $\mathbf{C}$ is an off diagonal matrix with zeros in the diagonal and coupling coefficient between each pair of cores in the off-diagonal elements. By diagonalizing the coupling matrix, 
\begin{equation}
    \mathbf{C} = \mathbf{V} \mathbf{B} {\mathbf{V}}^{T}
\end{equation}
a new set of modes is defined which is the supermodes i.e the perturbed eigenmodes, $\mathbf{V}$, with their propagation constants $\Delta{\beta}^{(j)}$ given as the eigenvalues i.e. the diagonal elements of $\mathbf{B}$.
The effective indices can be found as: 
\begin{equation}
    n_{\mathrm{eff}}^{(j)} = n_{\mathrm{eff,0}} + \frac{\lambda}{2\pi} \Delta \beta^{(j)}
\end{equation}
with $n_{\mathrm{eff,0}}$ the effective index of the unperturbed fundamental mode.
Consequently, we are able to compute the transmission matrix (TM) $\mathbf{H}$ of the fiber in the supermode basis with the element-wise exponential $\mathrm{exp} {(i \mathbf{B} L)} $ where $L$ is the Length of the MCF. We can finally transform this supermode TM into the basis of core modes as
\begin{equation}
    \mathbf{H} = \mathbf{V} \mathrm{e}^{i \mathbf{B} L} {\mathbf{V}}^{T} .
\end{equation}
The TM is a numerical representation of the fiber, thus it provides a convenient characterization of the fiber. $\mathbf{H}$ is $N\times N$ matrix ie expressed in the basis of the unperturbed fundamental modes of the cores. Generally speaking the modes that have close propagation constants are likely to exchange energy during propagation. Conversely, if the difference in propagation constant of modes is large, the likelihood of energy exchange is reduced. It was established that the TM of a MCF with uncoupled cores is a diagonal matrix. Inter-core coupling in MCF results in the appearance of off-diagonal elements in the TM.

The principal property that can be derived from $\mathbf{H}$ alone is the cross-talk (XT). Generally, the XT is defined as the relative amount of energy exchanged between cores. In our model, we would like to define the XT matrix $\mathbf{X}$ as the average of the norm-square of the $\mathbf{H}$ generated with different wavelengths $\lambda$, and so the average over $\lambda$ is the average over a sufficiently large number of calculated TMs for different wavelengths. Consequently, we obtain a $N \times N$ "cross-talk matrix" represented by the following equation:

\begin{equation}
  \mathbf{X} = \langle | \mathbf{H} |^{2} \rangle _{\lambda}
\end{equation}

In case of uncoupled cores, $\mathbf{X}$ is a diagonal matrix with $X_{ii}$~=~0~dB. This means that beam of light injected into a single core $i$ in the proximal end remains guided by core $i$ during propagation. However, when inter-core coupling exist, $\mathbf{X}$ becomes an off-diagonal matrix with diagonal elements $X_{ii}$ representing the amount of energy remaining in the excited core $i$, and the off-diagonal elements $X_{ij}$ representing the amount of energy transferred from excited core $i$ to coupling core $j$ during propagation. Therefore, the average amount of coupling between MCF cores is represented by the scalar "average total cross-talk" $X_{\mathrm{ave}}$ which is calculated by the following equation:

\begin{equation}
  X_{\mathrm{ave}} = \frac{1}{N} \sum_{i} (\sum_{j \ne i} X_{ij} )
  \label{SI:eq:Xave}
\end{equation}

The CMT model also returns the field of the fundamental core mode vs the radial coordinate; it can be interpolated on the cartesian "distal pixel basis" $\{ x^{(\mathrm{dist})}_j,y^{(\mathrm{dist})}_j \}, 1 \le j \le N_{\mathrm{pix}}$; centering $N$ copies on the core positions $(x^{(\mathrm{core})}_{j}, y^{(\mathrm{core})}_{j}), 1 \le j \le N$ ; each of these can be numerically propagated to the far-field by a two-dimensional discrete Fourier transform, the resulting fields are expressed in the (angular) "far-field pixel basis" $\{ f^{(\mathrm{FF})}_{x,j},f^{(\mathrm{FF})}_{y,j} \}, 1 \le j \le N_{\mathrm{pix}}$; and finally re-arranged to a mode-to-far-field pixel basis change matrix $\mathbf{U}$ of dimensions $N \times N_{\mathrm{pix}}$. 
$\mathbf{U}$ together with $\mathbf{H}$ permit to calculate ${\mathbf{e}}^{(\mathrm{FF, pix})}$ of dimensions $N_{\mathrm{pix}} \times 1$ the field in the far-field of the MCF distal end face
\begin{equation}
  {\mathbf{e}}^{(\mathrm{FF, pix})} =  {\mathbf{U}}^{\dagger} \mathbf{H} {\mathbf{e}}^{(\mathrm{prox, mode})}
\end{equation}
where ${\mathbf{e}}^{(\mathrm{prox, mode})}$ is $N \times 1$ and contains elements representing the complex modal field amplitudes injected into the proximal end of the MCF. It is from this vector that the properties of Strehl ratio (a measure of the intensity in the focus) and the memory effect (a measure of input-output angular correlations allowing distal point-scanning by proximal angular scanning) \cite{freund1988memory} can be inferred. 
${\mathbf{e}}^{(\mathrm{FF, pix})}$ can also be numerically propagated to any intermediate plane with an angular spectrum propagator if required.
The elements of the $N \times 1$ vector ${\mathbf{e}}^{(\mathrm{prox},\mathrm{mode},j_{0})}$ are the optimal input modal amplitudes to obtain an optimal focus on far-field pixel $j_{0}$:
\begin{equation}
  {\mathbf{e}}^{(\mathrm{prox},\mathrm{mode},j_{0})} : {e}^{(\mathrm{prox},\mathrm{mode},j_{0})}_k = ( A_{jk} )^{*}, j = j_{0}, 1 \leq k \leq N
\end{equation}
with
\begin{equation}
    \mathbf{A} = \mathbf{U}^{\dagger}\mathbf{H}.
\end{equation}
${\mathbf{e}}^{(\mathrm{FF, pix, j_{0}})}$ is the resulting complex field optimally focused on far-field pixel $j_{0}$: 
\begin{equation}
  {\mathbf{e}}^{(\mathrm{FF, pix, j_{0}})} = {\mathbf{U}}^{\dagger} \mathbf{H} {\mathbf{e}}^{(\mathrm{prox},\mathrm{mode},j_{0})} .
  \label{SI:eq:EFFpix}
\end{equation}
The Strehl ratio $S$ can then be defined as the ratio of the energy in the focus on far-field pixel $j_{0}$ to the total energy in the field ($P$ is the set of far-field pixels within one 1/$e^{4}$ half-width of far-field pixel $j_{0}$---where intensity is maximum):
\begin{equation}
  S = \frac{ \sum_{k \in P } | e^{(\mathrm{FF, pix, j_{0}})}_{k} |^{2}  }
  {\sum_{k = 1}^{N_{\mathrm{pix}}} | e^{(\mathrm{FF, pix, j_{0}})}_{k} |^{2} }
  \label{SI:eq:Strehl}
\end{equation}
The so-called "memory effect" is the effect by which the direction (phase ramp) of the output field changes by the same amount as the direction (phase ramp) of an input field \cite{freund1988memory}. In presence of the memory effect, it is therefore possible to displace a distal focus by simply adding a phase ramp to the proximal field.
${\mathbf{p}}(f^{(0)}_{x}, f^{(0)}_{y})$ is $N \times 1$ and its elements are the [complex exponential of] the additional proximal modal phases required to move the focus in the far-field by $(f^{(0)}_{x},f^{(0)}_{y})$ using the memory effect:
\begin{equation}
  \mathbf{p}(f^{(0)}_{x},f^{(0)}_{y}) : p_{k}(f^{(0)}_{x},f^{(0)}_{y}) = \mathrm{exp} (i 2 \pi f^{(0)}_{x} x_{k}^{(\mathrm{core})} + i 2 \pi f^{(0)}_{y} y_{k}^{(\mathrm{core})} )
  \label{SI:eq:phaseramp}
\end{equation}
The memory effect can then be quantified as a two-dimensional curve, the denominator contains the intensity of the field "optimally focused on far-field pixel $j_{1}$" cf Eq.~(\ref{SI:eq:EFFpix}); and the numerator contains the intensity of the field "initially optimally focused on far-field pixel $j_{0}$ then shifted to far-field pixel $j_{1}$ using the memory effect":
\begin{equation}
  M(f_{x}, f_{y}) = \frac{ | 
  e^{(\mathrm{FF, pix}, j_{0} \rightarrow j_{1})}_{j_{1}}
  |^2 }
  { | e^{(\mathrm{FF, pix}, j_{1})}_{j_{1}} |^{2} }
  \label{SI:eq:M}
\end{equation}
where "$\circ$" signifies Hadamard product (element-by-element multiplication) and
\begin{equation}
  {\mathbf{e}}^{(\mathrm{FF, pix}, j_{0}\rightarrow j_{1})} = {\mathbf{U}}^{\dagger} \mathbf{H} [ {\mathbf{e}}^{(\mathrm{prox},\mathrm{mode},j_{0})} \circ {\mathbf{p}}(f^{(0)}_{x}, f^{(0)}_{y}) ] 
\end{equation}
$M$ is constant =~1 for "full" memory effect, and =~0 everywhere expect the origin for "no" memory effect. 

\begin{table}[htbp]
\centering
\begin{tabular}{|c|c|l|}
    \hline
    \textbf{Variable} & \textbf{Input/Output} & \textbf{Description} \\
    \hline
    \verb|Lseg_array| & Input & \begin{minipage}{7cm}
        \vspace{0.2cm}
        [1$\times N_{\mathrm{seg}}$]. Units meters.
        Contains the lengths of each of the $N_{\mathrm{seg}}$ segments into which the simulated tapered MCF is sub-divided.
        \vspace{0.2cm}
        \end{minipage}\\
    \verb|lambda| & Input & \begin{minipage}{7cm}
        Scalar. Units meters. The wavelength of the laser. 
        \vspace{0.2cm}
        \end{minipage}\\
    \verb|a0| & Input & \begin{minipage}{7cm}
        Scalar. Units meters. Radius of a single core in a segment where \verb|t_array = 1|. Article equivalent $a$, the radius of a single core in the transport segment.
        \vspace{0.2cm}
        \end{minipage}\\
    \verb|n1| & Input & \begin{minipage}{7cm}
        Scalar. Unitless. The maximum of the refractive index profile of a single core. 
        \vspace{0.2cm}
        \end{minipage}\\
    \verb|n2| & Input & \begin{minipage}{6cm}
        Scalar. Unitless. The cladding refractive index.
        \vspace{0.2cm}
        \end{minipage}\\
    \verb|poscore0| & Input & \begin{minipage}{7cm}
        [$N\times$2]. Units metres. Positions of the $N$ cores of the MCF in a segment where \verb|t_array = 1|. 
        Article equivalent \{$x_{j}^{(\mathrm{core})}$,$y_{j}^{(\mathrm{core})}$\}.
        \vspace{0.2cm}
        \end{minipage}\\
    \verb|t_array| & Input & \begin{minipage}{7cm}
        [1$\times N_{\mathrm{seg}}$]. Unitless. Contains the homothetic scaling factors of each of the  $N_{\mathrm{seg}}$ segments into which the simulated tapered MCF is sub-divided. Ie in a given segment the core radius and the core positions are scaled by this factor.
        \vspace{0.2cm}
        \end{minipage}\\
    \verb|neff| & Output & \begin{minipage}{7cm}
        Scalar. Unitless. The effective index of the fundamental mode of an unperturbed core, ie without accounting for interaction between cores.
        \vspace{0.2cm}
        \end{minipage}\\
    \verb|E| & Output & \begin{minipage}{7cm} 
        [$N\times$1]. Arbitrary units. The field of the fundamental mode of an unperturbed core as a function of the radial coordinate contained in variable \verb|r|. 
        \vspace{0.2cm}
        \end{minipage}\\
    \verb|r| & Output & \begin{minipage}{7cm}
        [1$\times N$]. Units meters. The radial coordinates at which variable \verb|E| is evaluated.
        \vspace{0.2cm}
        \end{minipage}\\
    \verb|neffsuper| & Output & \begin{minipage}{7cm}
        [$N\times N_{\mathrm{seg}}$]. Unitless. The effective indices of the $N$ supermodes of the tapered MCF, ie of the perturbed eigenmodes found by CMT. Contains $N_\mathrm{seg}$ columns correponding to the segments into which the simulated tapered MCF is sub-divided.
        \vspace{0.2cm}
        \end{minipage}\\
    \verb|TM| & Output & \begin{minipage}{7cm}
        [$N\times N$]. Unitless. TM of the simulated tapered MCF in the basis of the fundamental modes of the $N$ cores. Article equivalent $\textbf{H}$.
        \vspace{0.2cm}
        \end{minipage}\\
    \hline
\end{tabular}
\caption{Overview of the Matlab function \texttt{CMT\_function.m} available in Dataset 1, Ref.~\cite{Dataset1}}.
\label{SI:tab:variables}
\end{table}

\section{Model based on finite-element beam propagation method}
\label{SI:sec:Finite_element_model}

To accurately simulate the propagation of light in longitudinally
varying waveguides, we used a homemade scalar wide-angle beam
propagation method \cite{koshiba1996wide} based on a finite element method (FE-BPM)
with first order elements.

We first designed the MCF cross-section in the transverse $x$–$y$ plane by
creating a non-uniform mesh of triangular elements. The field $E_{0}$ of
the fundamental mode in one core was computed with a 1D-finite
difference method and was interpolated on the structure of the MCF to
obtain the $E_{\mathrm{initial}}$ matrix containing the fundamental mode
field of one of the $N$ cores of the MCF. Then we use our FE-BPM
algorithm to propagate the field of each core individually along the
$z$-direction with a step of 1~$\upmu$m, and thus generate the
$E_{\mathrm{final}}$ matrix at each propagation step. This is repeated for 
$E_{\mathrm{initial}}$ centered on each of the $N$ cores.
We compute the TM $\mathbf{H}$ by projecting the $N$ resulting
$E_{\mathrm{final}}$ on the non-orthogonal basis formed by the 
fundamental modes $E_{0}$ of the $N$ cores, obtaining a $N\times N$ matrix in the
core (fundamental mode) basis. When
$E_{\mathrm{final}}~=~E_{\mathrm{initial}}$ for all cores, this
means that core fields do not loose any energy during propagation
and thus XT in the MCF is zero, and
$\mathbf{H}$ is diagonal. However, when XT is non-zero, $E_{\mathrm{final}}$ 
differs from $E_{\mathrm{initial}}$ and thus $\mathbf{H}$ will contain some 
non-zero off-diagonal elements. Consequently, from $\mathbf{H}$ alone we 
are able to compute $\mathbf{X}$ by the same
way mentioned in Suppl. Doc. Sec.~\ref{SI:sec:Coupled_mode_theory_model}.

\section{Methods: Fiber drawing}
\label{SI:sec:Methods_fiber_drawing}

The fiber has been made using the stack and draw process \cite{russell2006photonic}. A germanium doped silica preform with a parabolic index profile (maximal index contrast of 30$\cdot$10$^{-3}$ compared to its silica cladding) has been drawn into hundreds of rods of 1.46~mm diameter with a doped diameter of about 0.52 mm. These rods were then stacked manually on a hexagonal lattice and jacketed into a silica tube of 34~mm outside diameter. This assembly containing the 349 cores was then drawn at about 2000~$^{\circ}$C into canes of about 5~mm while applying a vacuum inside the jacketing tube to collapse the interstitial holes between the rods constituting the stack. Finally, one of these canes was drawn into a 330~$\upmu$m fiber at 1990~$^{\circ}$C with relatively slow speed (feed rate of 5~mm/min for a drawing speed of 12~m/min). 

\section{Inter-core group-delay dispersion}
\label{SI:sec:IntercoreGDD}
This question has been treated previously in Refs.~\cite{roper2015minimizing, roper2015advances, andresen2015measurement}. 
If all cores were identical, their group indices (and hence group velocities) would be identical as well. It can be hypothesized that inter-core group delay dispersion arises from uncorrelated variations in core diameters along the length of the MCF. The inter-core group-delay dispersion is then directly proportional to the derivative of the group index with respect to core radius $\mathrm{d}n_{g}/\mathrm{d}a$. 
Variation can be hypothesized to appear during all fabrication steps: from the deposition of the doped silica during the preform fabrication up to the final fiber draw (see Sec.~\ref{SI:sec:Methods_fiber_drawing} for more details).
Even in the case of an ideal stack, its draw can lead to small transverse variations between cores from at least two possible origins: (i) uneven drawing speed; (ii) uneven heating. In case (i) core radii $a$ vary while the maximum core refractive indices $n_{1}$ are preserved; in case (ii) core radii $a$ vary while maximum core refractive indices vary as $a^{-2}$ as dopant quantity is preserved in a transverse plane. 
The impact of (i) and (ii) are predicted for four different initial index profiles with grading parameters $\alpha$~=~1, 2 (parabolic), 4, and $\infty$ (step). Effective indices of the fundamental mode are calculated by a scalar mode solver as a function of wavelength $\lambda$ and core radius $a$ with all other parameters similar to those given in the main article. 
Group indices are then calculated as
\begin{equation}
    n_{g} = n_{\mathrm{eff}} - \lambda \frac{\mathrm{d}n_{\mathrm{eff}}}{\mathrm{d}\lambda}
\end{equation}
from which the sought derivative $\mathrm{d}n_{g}/\mathrm{d}a$ is subsequently calculated numerically as 
\begin{equation}
    \frac{\mathrm{d}n_{g}}{\mathrm{d}a} \approx \frac{n_{g}(\lambda + \frac{1}{2}\Delta\lambda) - n_{g}(\lambda - \frac{1}{2}\Delta\lambda) }{\Delta \lambda}.
\end{equation}
\begin{figure}
    \centering
    \includegraphics[width=\textwidth]{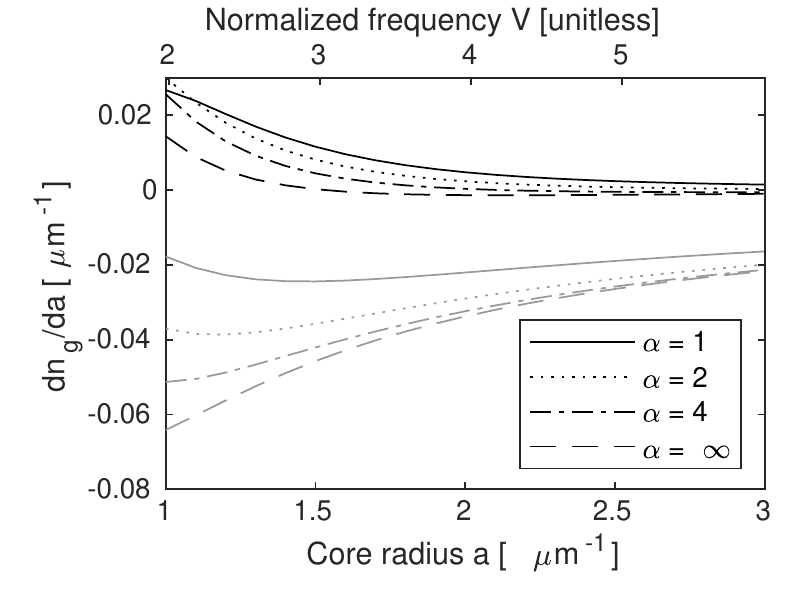}
    \caption{Calculated derivatives of the group delay vs. core radius for the four combinations of index profile (step and parabolic) and origin of core radius variation (black curves correspond to (i) uneven drawing speed and grey curves to (ii) uneven heating) as described in the text. The parameters of the calculation were taken similar to the experimental parameters in the main article.}
    \label{SI:fig:intercoreGDD}
\end{figure}
Calculated derivatives are presented in Fig.~\ref{SI:fig:intercoreGDD} for four different grading parameters. 
The results permit to conclude that a step-index profile ($\alpha$~=~$\infty$) eliminates inter-core group-delay dispersion ($\mathrm{d}n_{g}/\mathrm{d}a$~=~0 for $V$~=~3---dashed line) but under the condition that (i) is the physical origin. If (ii) is the origin the step index profile is the worst (dash-dotted line). 
In practice, it is known than dopant does diffuse during fiber drawing, effectively smoothing refractive index profiles in the process. This is particularly marked for the small core diameters relevant for the present study \cite{yablon2005optical}. It is then to be expected that an initial step-index profile in the pre-form evolves toward a gaussian profile during drawing. 
The dotted and full lines in Fig.~\ref{SI:fig:intercoreGDD} show curves for case (i) and (ii) for parabolic index profile (a parabolic profile can be thought of a second-order approximation of a gaussian profile). While none of these curves exhibit a zero, they do converge uniformly toward zero for increasing core radius. 
According to this example, the trend is that inter-core group-delay dispersion (of the fundamental mode) is reduced with bigger core radius.

\section{Methods: Fiber post-processing (tapering)}
\label{SI:sec:Methods_fiber_post_processing}

\begin{figure}[htbp]
	\centering
	\includegraphics[width=\textwidth]{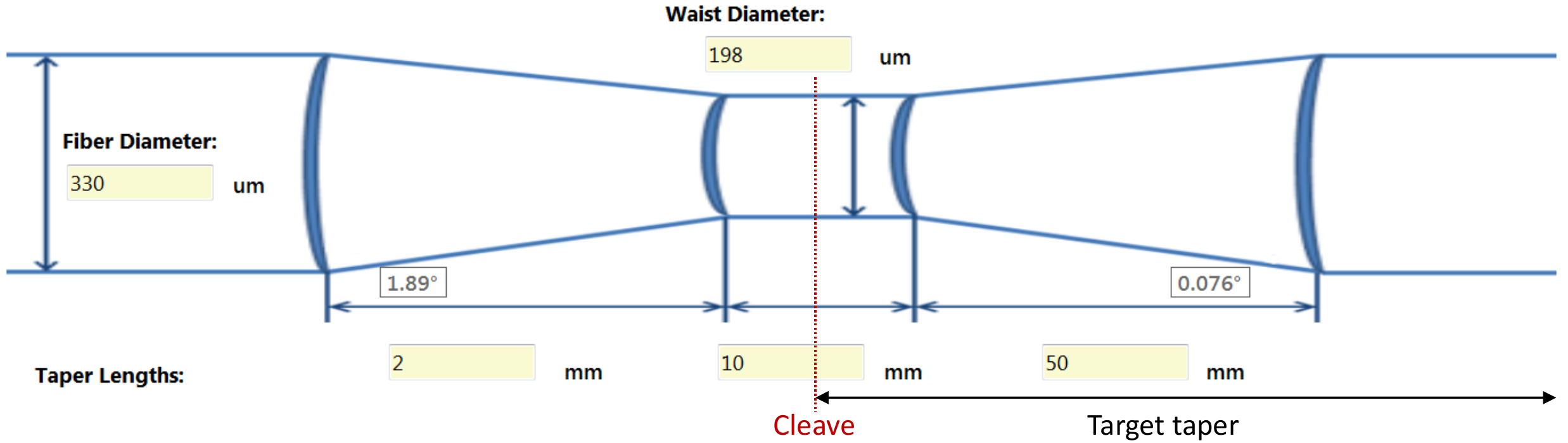}
	\caption{Screenshot from the fiber processing software showing the target profile and the parameters that can be entered. The dashed vertical line indicated the point where the bi-conical taper is cleaved after processing to yield one extremity (either the injection or the imaging segment) of the final tapered MCF. The given example is for a taper ratio of 0.6.}
	\label{SI:fig:Physicalcharacteristics}
\end{figure}

The fabrication presented in Suppl. Doc. Sec.~\ref{SI:sec:Methods_fiber_drawing} produced tens of meters of MCF.
In the post-processing described here, the MCF was divided into shorter lengths, whose two ends were tapered using a CO$_{2}$ laser-based glass processing and splicing system (LZM-100 FUJIKURA) which provides extremely stable operation that allows control over a wide range of parameters. Noting that, in order to end up with a 30~cm tapered fibers, we pre-cut a 30 + 2~$\cdot$~(15)~cm from the spool of the  fabricated MCF, as a typical length. 

The tapering process is achieved by a combined local heating of the fiber (by CO$_{2}$ laser beam) and simultaneously applying a pulling force lengthwise on both ends of the fiber. Once the tapering process ended, we obtain a bi-conical tapered optical fiber connected with the waist, similar to Fig.~\ref{SI:fig:Physicalcharacteristics}. Then we remove the fiber carefully, and we cleave it using CT-106 FUJIKURA cleaver in the middle of the waist (dashed line in Fig.~\ref{SI:fig:Physicalcharacteristics}), and thus we obtain two pieces, one long piece of an optical fiber with adiabatic taper at one end, and this is the target taper. And the second (15~cm) whose cleaved end is measured under a microscope to check whether the cleave is clean after which it is discarded. Then we repeat the same process on the un-tapered side of the realized tapered fiber, and thus to obtain a double tapered MCF as shown in Fig.~1 in the main article.

Specific details of the process are given in the following. The parameters of the target profile, a bi-conical taper, is specified in the fiber processing software which controls the LZM-100. Fig.~\ref{SI:fig:Physicalcharacteristics} is a screenshot from the software giving an example of such a profile and the parameters that can be set to define it. From this, the software automatically calculates the tension and pulling speed to be applied by the stepper motors of the LZM-100. 
Before the MCF is placed in the LZM-100 the polymer coating is completely removed over the length of the MCF, this is an essential step especially for the area that will be exposed to laser beam, using a razor blade and cleaned using a tissue soaked with isopropanol or ethanol. Then  fiber is fixed in the LZM-100 by clamping its terminals by fiber holders chosen with appropriate size to fibers diameter. These holders are based on the stepper motors that will apply the pulling force lengthwise on both ends of the fiber. Fibers are also fixed tightly using a split v-groove clamps  placed on the left and right of the laser source, these v-groove clamps are automatically adjusted to fiber diameter. 
The tapering process typically takes 200~s. When it is finished the software can measure the actual profile using the internal cameras of the LZM-100 and create a report in Excel format containing the target profile and the measured profile. We usually start with a test draw to check the suitability of automatically calibrated parameters. From data in the extracted report, if the curve of measured profile is above the target, then laser heating power is insufficient. If measured curve exhibits large ripple, then  laser heating power is excessive. In both cases, we should manually modify some of power parameters by increasing or decreasing their values by small percentage, until we identify the optimal power for obtaining the target profile. 

\section{Methods: Optical properties of fabricated tapered multi-core fibers}
\label{SI:sec:Methods_optical_properties_of_fabricated_tapered_multi_core_fibers}

\begin{figure}[htbp]
	\centering
	\includegraphics[width=\textwidth]{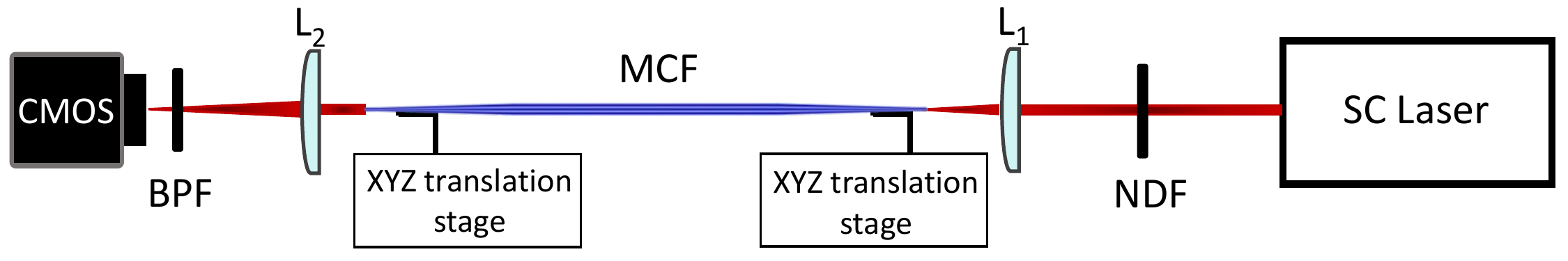}
	\caption{Experimental setup for measuring optical properties of tapered MCF. 
	SC Laser, super-continuum laser source (YSL Photonics SC-PRO); 
	NDF, neutral density filter; 
	L1, lens with focal length 4.05~mm; 
	MCF, tapered MCF; 
	L2, lens with focal length 4.51~mm; 
	BPF, band-pass filter (Thorlabs, FB900-10); 
	CMOS, CMOS camera (Thorlabs, DCC1545M).}
	\label{fig:setup_charac_tapered_MCFs}
\end{figure}

The experimental setup (Fig.~\ref{fig:setup_charac_tapered_MCFs}) was designed to measure the optical properties of tapered MCF using a super-continuum laser source (YSL Photonics SC-PRO). Both proximal and distal ends of the double tapered MCF are fixed on XYZ translation stages. The laser beam entering the proximal end of the MCF is focused by an aspheric lens L$_1$ with focal length 4.03~mm into one core at a time. We use the XYZ translation stage to move the focus from core to core. The emerging light at the distal end of the fiber is imaged with aspheric lens L$_2$  of 4.51~mm focal length onto a monochrome CMOS camera (Thorlabs, DCC1545M). A bandpass filter (FB900-10) is placed in front of the camera in order to measure the tapered MCF properties at a wavelength of 900~nm.

All tested tapered MCFs were tapered at the proximal end with $t^{(\mathrm{inj})}$~=~0.6 to assure injection into only the fundamental mode of each core.

In the absence of XT the recorded images are composed of a gaussian-like spot which corresponds to the intensity profile of the fundamental mode of the injected core. The mode field diameter (MFDs) of the fundamental mode is measured as the full-width of the intensity profile at $1/e^2$ of its maximum.

In the case of non-zero XT a number of additional spots show up in the recorded image. The XT matrix $\mathbf{X}$ is measured as follows.
The intensities $I_{j}$ of the spots are measured from the image. The $i$'th line of $\mathbf{X}$ is $X_{ij}~=~I_{j}/\sum_{j}I_{j}$ where $i$ is the number of the core one is injecting into. This is repeated for all cores $1 \le i \le N$ to construct the full XT matrix. 
In practice the full measurement of $\mathbf{X}$ is extremely tedious and time consuming. 
A measure of the average total XT $X_{\mathrm{ave}}$ is more useful for comparing with the predictions of the CMT model. It can be estimated by measuring $X_{ij}$ for a small number $N_{\mathrm{red}}$ of randomly chosen core numbers $i$ and then calculating by the appropriately modified Eq.~(\ref{SI:eq:Xave}).
\begin{equation}
  X_{\mathrm{ave}} = \frac{1}{N_{\mathrm{red}}} \sum_{i} (\sum_{j \ne i} X_{ij} ) .
\end{equation}

\begin{figure}[htbp]
	\centering
	\includegraphics[width=\textwidth]{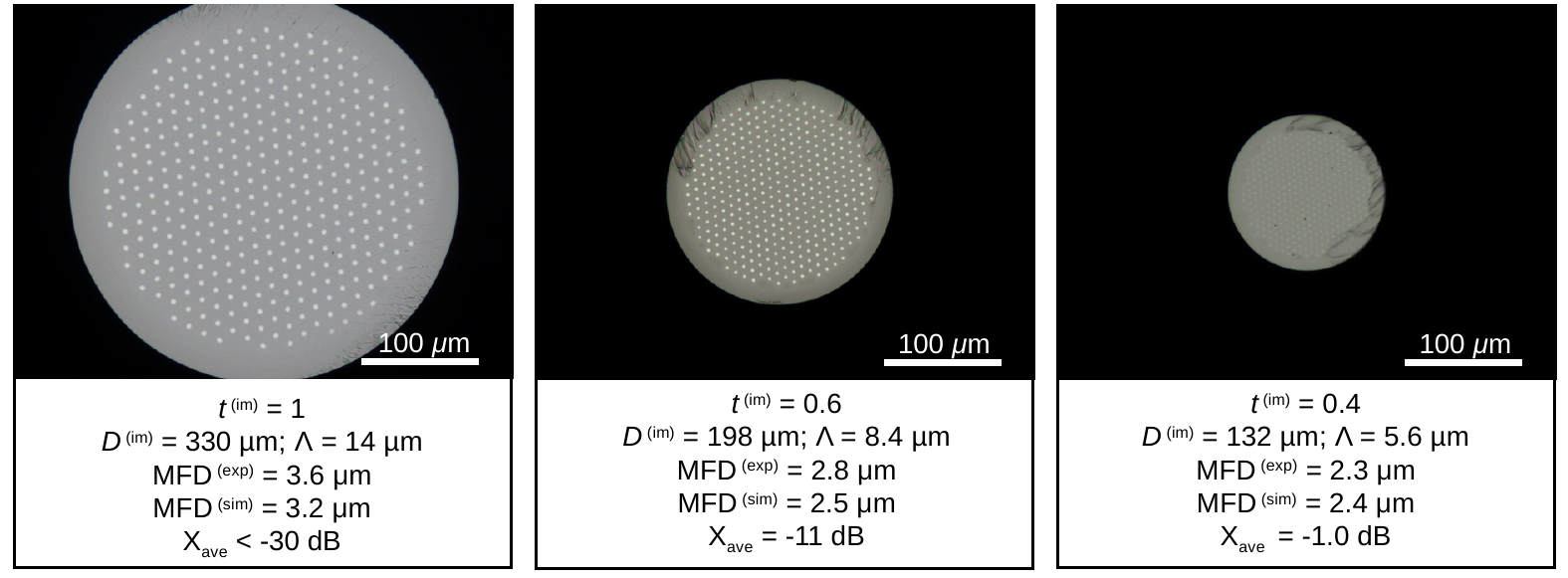}
	\caption{Distal end faces of the tapered MCFs from Tab.~1 in the main article with taper ratios $t^{(\mathrm{im})}$~=~1, 0.6 and 0.4. All images are on the same scale. $D^{(\mathrm{im})}$, diameter of imaging segment. $\Lambda$, pitch in the imaging segment. MFD$^{(\mathrm{exp})}$, experimentally measured mode field diameter [$\upmu$m]: $X_{\mathrm{ave}}$, measured average total XT [dB].  }
	\label{SI:fig:properties}
\end{figure}

Additionally, microscope images were acquired of the distal end faces of the fabricated tapered fibers ie their imaging segment. 
Fig.~\ref{SI:fig:properties} shows the microscope images of the distal end faces of the three tapered MCFs with parameters as in Tab.~1 in the main article as well as the measured values of their diameter $D^{(\mathrm{im})}$; pitch $\Lambda$; the MFDs MFD$^{(\mathrm{exp})}$; and average total cross-talk $X_{\mathrm{ave}}$. Reported MFD$^{(\mathrm{exp})}$ values represent an average of MFDs calculated for multiple cores. 

\section{Methods: Lensless endoscope}
\label{SI:sec:Methods:Lensless_endoscope}

\begin{figure}[htbp]
	\centering
	\includegraphics[width=\textwidth]{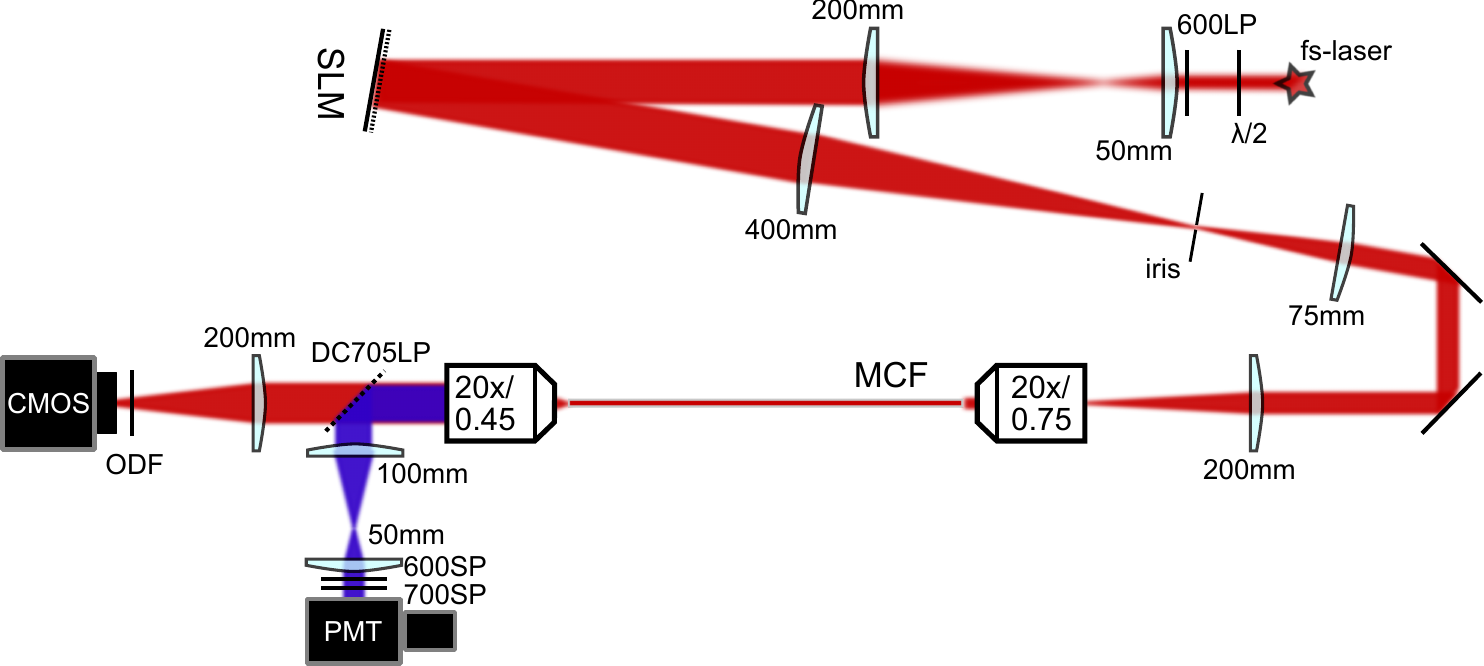}
	\caption{Lensless endoscope setup used to characterize the TM, memory effect, Strehl ratio etc. of the tapered MCFs and subsequently to perform focus scanning two-photon-imaging in the forward direction.
	CMOS, CMOS camera (FLIR, BFLY-U3-23S6M-C).
	ODF, neutral density filter.
	PMT, photo-multiplier tube (Hamamatsu, R9110).
	20$\times$/0.45, microscope objective (Olympus, LUCPlanFLN, 20$\times$ / 0.45~NA).
	MCF, tapered multi-core fiber.
	20$\times$/0.75, microscope objective (Nikon, CFI Plan Apo, 20$\times$ / NA~0.75).
	SLM, spatial-light modulator (Hamamatsu, X10468-07)
	\label{SI:fig:Setup_lensless_endoscope}}
\end{figure}

\begin{figure}
	\centering
	\includegraphics[width=\textwidth]{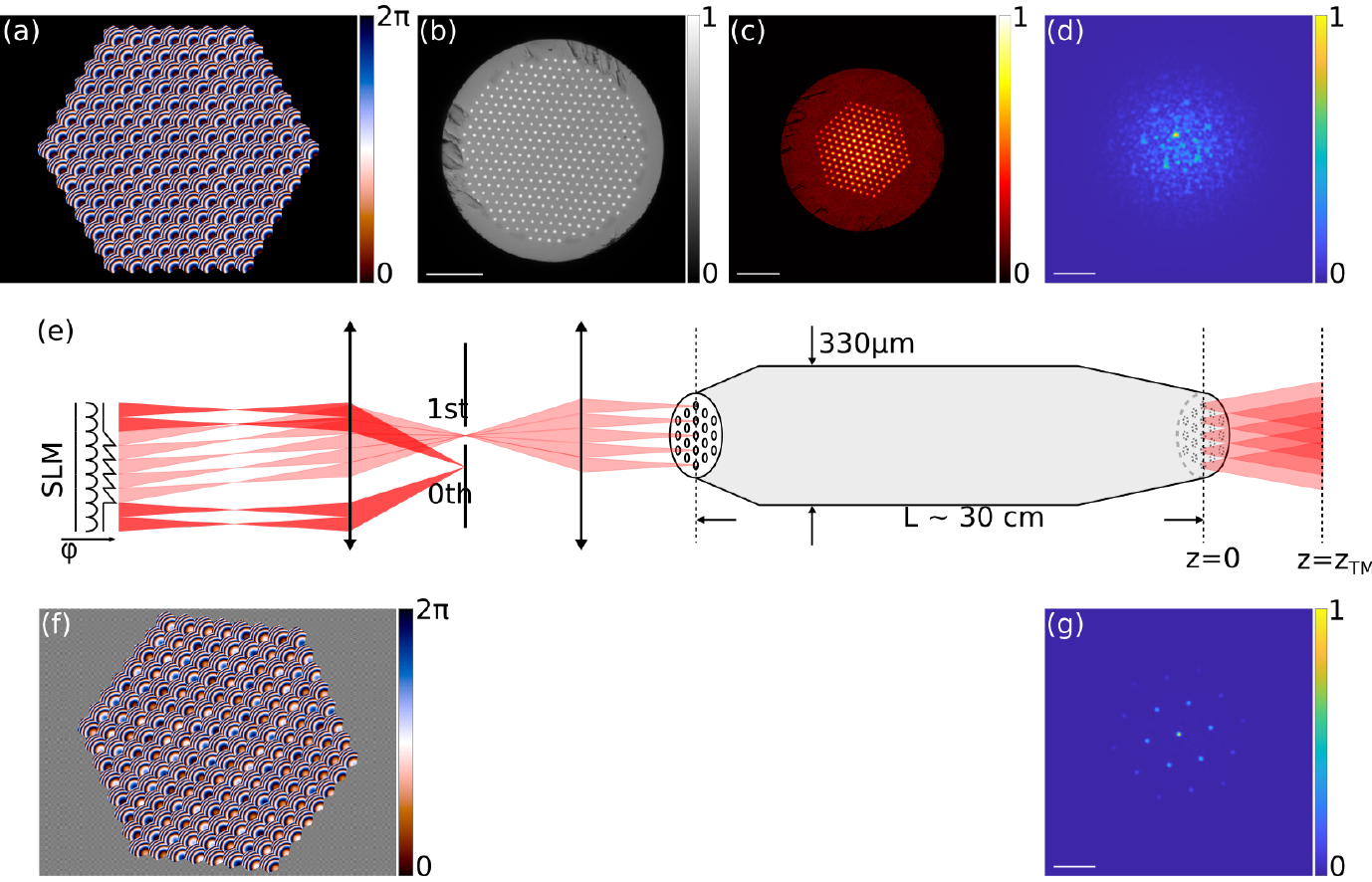}
	\caption{Schematic of incoupling of light at the proximal end and focusing at the distal end of a tapered MCF. Exemple for $t^{(\mathrm{im})}$~=~0.6.
	(a) Phase mask on SLM for incoupling into 169 cores. 
	(b) Microscope image of the MCF end face.
	(c) Intensity profile in the plane $Z$~=~0. 
	(d) Speckle seen in the plane $Z$ with the phase mask (a) displayed on the SLM. 
	(e) Sketch of light paths from SLM to the distal end of the tapered MCF.
	(f) Example of a phase mask which results in a focus at the distal end of the tapered MCF.
	(g) Example of intensity distribution in the plane $Z$ with the phase mask (f) displayed on the SLM.
	SLM, spatial-light modulator. Scale bars, 50~$\upmu$m.
	\label{SI:fig:principle_incoupling_focusing}}
\end{figure}

The experimental setup is sketched in Fig.\ref{SI:fig:Setup_lensless_endoscope}.
The experiments were conducted with a pulsed fs-laser (Chameleon Ultra II, Coherent) tuned to 920~nm wavelength. It has a repetition rate of 80~MHz and 150~fs pulse length. The laser beam is expanded by a keplerian telescope with a magnification of 4 to fill the active area of the spatial light modulator (SLM) (X10468-07, Hamamatsu). As in Ref.~\citenum{andresen2013two} the SLM is used for multiple purposes, mainly to phase the individual fiber cores to form a coherent focus for nonlinear imaging. However, it serves equally to create the micro-lens array for efficient light coupling into the individual fiber cores of the MCF [Figs.~\ref{SI:fig:principle_incoupling_focusing}(a), \ref{SI:fig:principle_incoupling_focusing}(e)], and is used as a selective blaze grating that allows to individually switch on fiber cores for the measurement of the TM. To ensure the unused parts of the SLM are not coupled in the fiber, an iris is used to cut off the 0th order of this blazed grating. The half-wave plate ($\lambda/$2) before the telescope rotates the linear polarization to match the orientation of the extraordinary axis of the SLM which is horizontal in the given setup. The combination of the following three lenses (400~mm, 75~mm and 200~mm) and the microscope objective (CFI Plan Apo, 20$\times$ / NA~0.75, Nikon) ensures to demagnify the hexagonal tiling from the SLM to the proximal end of the fibers. Thus, the system magnification is $M_{\mathrm{sys}}= 400/75 \cdot M_{\mathrm{obj}} = 400/75 \cdot 20 = 106.67$. This allows to space the SLM such that 169 cores of the MCF can be addressed for incoupling. In order to measure the TM of the MCF, the emerging light at the distal end of the fiber is imaged with an objective (LUCPlanFLN, 20$\times$ / 0.45~NA, Olympus) and the tube lens of 200~mm focal length onto a standard CMOS camera (BFLY-U3-23S6M-C, FLIR). Optical density filters (ODF) prevent the camera from saturation. For collection of the two-photon fluorescence (TPF) signal in forward direction, a dichroic 705 long-pass (FF705-Di01-25$\times$36, Semrock) was put after the collection objective. The back-focal plane of the objective is imaged on an analogue photomultiplier tube (R9110, Hamamatsu). A combination of two short-pass filters (600SP and 700SP, Thorlabs) block leakage radiation from the excitation laser. The long-pass filter (600LP, Thorlabs) before the SLM additionally hinders pump diode leakage radiation from the fs-laser to interfere with the TPF signal detection on the PMT.

The main article states that the TM of the tapered MCF was measured as part of the experiments. We must stress that we actually use TM as shorthand for a subset of the full TM. The full TM would contain four quadrants corresponding to the four combinations of input and output polarizations; and all elements would be complex ie comport an amplitude and a phase. The TM measured here is strictly speaking the "[complex exponential of the] phases of one quadrant of the full TM". 
In the experiment it is measured as follows.
One core (typically the center core) is designated as the reference. Using the SLM we inject into the reference core and one other core and the phase of the reference is stepped through 0 to 7$\pi$/4 in eight steps. The three resulting interferograms on the CMOS camera, conjugated with the plane $Z$ are recorded. This is repeated for all $N$ cores resulting in an image stack of 8$N$ images. 
From each group of three images the phase difference between the two cores on any of the $N_{\mathrm{pix}}$ pixels of the CMOS camera can be found by the methods of phase-stepping interferometry. This gives $N_{\mathrm{pix}}$ phase values for each of the $N$ cores. [The complex exponential of ] these phase values are the elements of the sought TM.
The resulting TM is then a $N_{\mathrm{pix}} \times N$ matrix in the basis of (input) core modes and (output) camera pixels.

To establish a distal focus as in Figs.~4(a)-4(c) in the main article the appropriate phase mask had to be generated displayed on the SLM.
The line of the previously measured TM corresponding to a central pixel on the CMOS camera was identified, the phase values extracted and used to set the phases of the corresponding segments to produce a phase mask typically resembling Fig.~\ref{SI:fig:principle_incoupling_focusing}(f). Displaying this mask on the SLM resulted in an intensity image on the CMOS camera typically resembling Fig.~\ref{SI:fig:principle_incoupling_focusing}(g).

The distal focus can be moved across the ($x$,$y$)-plane at the distance $Z$ from the distal end facet of the tapered MCF in two ways which we refer to as "TM scan" and "ME scan".
The TM scan method uses the knowledge of the previously measured TM to generate a dedicated phase mask for each ($x$,$y$)-position of the distal focus similar to what was done to establish a central focus in the preceding paragraph.
When these masks were displayed in succession on the SLM the distal focus was scanned across the ($x$,$y$)-plane.
The ME scan method takes its starting point in the phase mask that results in a central focus. If the tapered MCF exhibits some ME then the distal focus can be moved by adding a small phase increment on each segment of the phase mask. This increment should be proportional to the segment's $x$- and $y$-coordinate, ie a phase ramp. The resulting displacement of the spot is proportional to slope of the phase ramp, the exact dependence can be found from Eq.~(\ref{SI:eq:phaseramp}).

The intensity images in Figs.~4(a)-4(c) in the main article were measured on the CMOS camera conjugated to the planes $Z$~=~500, 300, 200~$\upmu$m while the SLM displayed a phase mask resulting in a distal central focus.
The Strehl ratios presented in Fig.~4(e) in the main article were extracted from these images as $S$~=~(sum of the intensities of pixels comprising the central focus)/(sum of the intensities of all pixels), similar to Eq.~(\ref{SI:eq:Strehl}).

The data underlying Fig.~4(d) in the main article was two image stacks acquired on the CMOS camera.
The first stack was acquired by using the TM scan method as follows.
The TM scan method was used to generate the phase masks resulting in a number of distal foci on a line along the $x$-direction. At each position, the image of the optimized focus was acquired by the CMOS camera.
The second stack was acquired by using the ME scan method as follows.
The first image of the ME scan stack was identical to the first image of the TM scan stack, with the focus on $x$~=~0. 
Subsequently the focus was moved to the same $x$-positions as for the TM scan stack by using the ME scan method, and for each position an image was acquired by the CMOS camera.
The memory effect curve $M(x)$ in Fig.~4(d) in the main article was extracted from the two image stacks in the following way.
$M(x)$~=~(intensity of focus on $x$ produced by ME scan)/(intensity of focus on $x$ produced by TM scan) similar to Eq.~(\ref{SI:eq:M}).

The TPEF images in Figs.~5(a)-5(c) in the main article were acquired as follows.
The TM was measured with the CMOS camera conjugated to the plane $Z$.
Then the sample of fluorescent beads was placed in the plane $Z$, the focus was scanned across 60$\times$60 points in the ($x$,$y$)-plane by the TM method, and at each focus position the fluorescence signal was acquired by the PMT.

The TPEF images in Figs.~5(d)-5(e) in the main article were acquired as follows.
The TM was measured with the CMOS camera conjugated to the plane $Z$.
Then the sample of fluorescent beads was placed in the plane $Z$, the focus was scanned across 60$\times$60 points in the ($x$,$y$)-plane by the ME method, and at each focus position the fluorescence signal was acquired by the PMT.

The TPEF intensities appearing in Fig.~5(f) in the main article were extracted from the TPEF images by measuring the intensities of single, isolated beads.